\documentclass[a4paper,11pt]{article}
\usepackage{jheppub} 
\usepackage[T1]{fontenc}
\usepackage{amssymb, amsmath}
\usepackage{physics}
\usepackage{xcolor}
\usepackage{float}
\usepackage{graphicx}
\usepackage{subfigure}
\usepackage{subcaption}
\usepackage{natbib}
\usepackage{adjustbox}
\usepackage{tikz}
\usepackage{soul}
\usetikzlibrary{shapes,arrows}
\usepackage[compat=1.1.0]{tikz-feynman}
\tikzset{every picture/.style={/tikzfeynman/momentum/arrow distance=2mm}}
\usepackage[colorlinks=true,citecolor=blue,urlcolor=blue]{hyperref}
\usepackage{slashed}
\usetikzlibrary{shapes.geometric, arrows}
\usepackage{bbold}
\usepackage{slashed}
\usepackage{array}
\usepackage{comment}

\usepackage{multirow}
\usepackage{threeparttable}
\usepackage{paralist}
\usepackage{xspace}
\usepackage{upgreek}
\usepackage{lipsum}
\usepackage{mathrsfs}
\usepackage{feynmp-auto}

\expandafter\def\expandafter\normalsize\expandafter{%
    \normalsize%
    \setlength\abovedisplayskip{5pt}%
    \setlength\belowdisplayskip{5pt}%
    \setlength\abovedisplayshortskip{5pt}%
    \setlength\belowdisplayshortskip{5pt}%
}
\newcommand\colvec[3][]{\begin{pmatrix}\ifx\relax#1\relax\else#1\\\fi#2\\#3\end{pmatrix}}

\definecolor{darkmagenta}{rgb}{0.55, 0.0, 0.55}
\newcommand{\beq}{\begin{equation}}
\newcommand{\beqn}{\begin{eqnarray}}
\newcommand{\eeq}{\end{equation}}
\newcommand{\eeqn}{\end{eqnarray}}
\DeclareRobustCommand{\Eq}[1]{Eq.~\eqref{#1}}

\DeclareRobustCommand{\Sec}[1]{Section~\ref{#1}}
\DeclareRobustCommand{\App}[1]{Appendix~\ref{#1}}
\DeclareRobustCommand{\Fig}[1]{Fig.~\ref{#1}}

\newcommand{\deltabar}{\mathchar'26\mkern-9mu \delta}
\newcommand{\dkbar}[2][3]{\text{\dj}^{#1} #2\ }
\newcommand{\Disc}[1]{\text{Disc}\ #1}

\newcommand{\M}{\mathcal{M}}
\newcommand{\alphac}{{c}}
\newcommand{\ch}[1]{c_{#1}}
\newcommand{\chfull}[1]{\frac{1}{2}\coth(\frac{#1}{2T})}

\newcounter{step}
\newenvironment{step}{\refstepcounter{step} \textbf{Step~\arabic{step}. }}{}

\title{Spectral Surgery in a Heat Bath: a finite-temperature guide {to particle production} for phenomenologists}
\author{Nirmalya Brahma, Saniya Heeba, Hugo Sch\'erer, and Katelin Schutz}
\affiliation{Department of Physics \& Trottier Space Institute, 
McGill University, Montr\'{e}al, Canada}

\emailAdd{nirmalya.brahma@mail.mcgill.ca}
\emailAdd{saniya.heeba@mcgill.ca}
\emailAdd{hugo.scherer@mail.mcgill.ca}
\emailAdd{katelin.schutz@mcgill.ca}
\abstract{Quantifying the effects of finite temperature and density (FTD) on particle properties is essential for understanding phenomena within and beyond the Standard Model. In this work, we present a simplified framework for calculating particle production rates at FTD without resorting to a full thermal field theory calculation. We do so by relating the imaginary part of a particle's $n$-loop finite temperature self energy, which defines its in-medium damping rate, to a sum of thermally weighted tree-level vacuum rates. Such a mapping results in novel ``interference'' contributions to particle production which have no vacuum analog and which have been relatively overlooked in the phenomenology literature. These interference terms are known to regulate collinear and infrared divergences that arise when calculating interaction rates in a medium. We demonstrate the impact of these corrections with two toy models and find that properly accounting for these interference terms can alter particle production by an $\mathcal{O}(1)$ amount. We additionally compare the size of these corrections to the thermal mass corrections often studied in the literature, finding the sizes of these contributions to be of similar order. 
}

\begin{document}
\maketitle
\flushbottom

\section{Introduction}
It is well known that the properties of particles are significantly affected by their ambient environment~\cite{Das:1997gg, Kapusta:2006pm, Bellac:2011kqa}. In fact, the inclusion of finite temperature and density (FTD) effects has been crucial for accurately modeling Standard Model (SM) processes both in the early universe and in the evolution of astrophysical systems. For instance, accurate predictions of the abundance of light elements in the early universe, the temperature of neutrino decoupling, the physics of phase transitions, and stellar neutrino emission rely crucially on finite temperature calculations~\cite{Cielo:2023bqp,Esposito:1999sz,PhysRevD.46.3372,Mazumdar:2018dfl, Raffelt:1996wa}. Additionally, beyond the SM (BSM) phenomenology often relies on the interactions of weakly coupled particles in a variety of plasmas. For example, a thermal history of dark matter (DM) consistent with its observed abundance typically requires its production from the primordial SM plasma. Similarly, searches for well-motivated BSM particles such as dark photons and axions either rely on stellar plasmas as the source of these particles which can then be detected in the laboratory, or seek astrophysical signatures of in-medium BSM particle interactions~\cite{Raffelt:1996wa,Caputo:2024oqc,Carenza:2024ehj,Davidson:1991si,Davidson:1993sj,Hardy:2016kme,An:2020bxd,An:2013yfc,Viaux:2013hca,Viaux:2013lha,Isern:2008nt,Corsico:2012ki,Vogel:2013raa}. Consequently, understanding and quantifying particle propagation and interaction at FTD is foundational to BSM phenomenology, particularly in astrophysical and cosmological contexts. 

In recent years, there has been a growing interest in incorporating FTD effects more accurately in BSM phenomenology. In particular, freeze-in production of DM has been shown to be highly sensitive to the background, giving model-dependent corrections to the abundance that can be $\mathcal{O}(1)$ or even larger~\cite{Dvorkin:2019zdi,Bringmann:2021sth,Heeba:2019jho,Heeba:2018wtf,Becker:2023vwd}. FTD effects have also been crucial in studying BSM phenomenology in astrophysical systems, resulting in some of the strongest bounds on a range of BSM candidates such as axions, dark photons, light millicharged fermions, and scalars \cite{Fung:2023euv, Springmann_2025,Dolan:2023cjs,Hardy:2016kme}. The prevalence of these effects over a range of energy scales and ambient environments necessitates the development of a cohesive framework that can be straightforwardly applied to any system of interest to extract novel BSM phenomenology. 

FTD effects are generally calculated using tools from Finite Temperature Field Theory (FTFT). In particular, the modification to a particle's properties in the presence of a background can be encapsulated in its finite-temperature self-energy, $\Pi(\omega, \vec{k}) = \mathrm{Re}\Pi + i\,\mathrm{Im} \Pi$. The real part of the self-energy, $\mathrm{Re}\Pi=\omega^2-|\vec{k}|^2$ determines the particle's in-medium dispersion relation and therefore its effective mass, whereas the imaginary part, $\mathrm{Im}\Pi = -\omega \Gamma$ determines the interaction rate (discussed further in Sec. \ref{sec:prod_rate}). Therefore, the problem of quantifying the effects of an ambient environment is intimately linked to calculating the real and imaginary parts of $\Pi$ for the particle of interest using FTFT. 

Evaluating the particle self-energy in a medium requires calculating loop diagrams using FTFT, which can be unwieldy and numerically intractable. It was shown by Weldon that the imaginary part of a particle's in-medium self-energy, or its interaction rate, is equivalent to the particle interaction rate in vacuum with a phase space that is weighted by the thermal distribution functions of the particles in the background \cite{Weldon:1983jn}. This approach is widely used to calculate FTD rates as a proxy for the calculation using the full FTFT framework. However, this prescription does not include several important effects and is therefore incomplete. For instance, certain processes exist in a medium that do not have a vacuum analog. The most well-known of these is the decay of in-medium photons or ``plasmons'' into lighter SM or BSM particles \cite{PhysRev.129.1383,Dvorkin:2019zdi}. Furthermore, significant quantitative differences arise when computing BSM particle production rates using the procedure outlined above versus a full FTFT treatment \cite{Becker:2023vwd}. At a qualitative level, there can be important corrections to vacuum processes arising from the interference between $n\to m$ and $n+\ell \to m+\ell$ processes \cite{Wong:2000hq, Kapusta:2001jw}. These ``interference'' type contributions, which have been relatively overlooked in the BSM phenomenology literature, are not only necessary to regulate IR and collinear divergences arising at FTD~\cite{Majumder:2001iy}, but also result in a significant modification of the particle production rate. The omission of these effects indicates that the formalism often used in the literature is not complete, specifically for treating FTD effects when going beyond one-loop order in the particle self-energy. These higher-order loops in the self-energy correspond to interactions involving higher particle multiplicity. For instance, despite containing more powers of a small coupling constant, there are many situations where $2\to 2$ processes can dominate over $2 \leftrightarrow 1$ processes due to the kinematics or the availability of different species in the initial state. In these situations, obtaining the dominant thermal interaction rate would require computing a two-loop self-energy diagram. In general, when both $n\to m$ and $n+\ell\to m+\ell$ processes are significant sources of particle production, the formalism often used in the literature will not provide accurate rates. 

In this work, we present a set of rules to accurately quantify particle production rates at FTD, with an eye towards the accessibility of using these rules. In Section~\ref{sec:CutWeldon}, we review the usual approach in the literature linking $\mathrm{Im}\Pi$ to tree-level rates, and show how this approach fails at higher-loop order, in contrast to what is typically assumed. In Section~\ref{sec:SSHBrules}, we present our prescription to quantify FTD effects by interpreting the imaginary part of the full multi-loop self-energy in terms of tree-level processes. We discuss how to parameterize and interpret the novel interference corrections to these processes arising at multi-loop order. We corroborate the rules presented in this work by calculating the 2-loop self-energy of a toy model using ITF in Sec.~\ref{sec:corroborate}. Finally, in Section~\ref{sec:application}, we use our prescription to quantify the FTD impact on the particle production rate for two toy models. Concluding remarks follow in Section~\ref{sec:conclusion}.

\section{Calculating Observables at Finite Temperature}
\label{sec:CutWeldon}
\subsection{Particle production in a medium}
\label{sec:prod_rate}

At finite temperature and density, the presence of a medium modifies the dispersion and damping of particles compared to their behavior in a vacuum. These effects are encapsulated in the self-energy $\Pi$. The complex dispersion relation is
\begin{equation}
    \omega_c^2 - \vec{k}^2 - m^2 - \Pi = 0,
\end{equation}
where $\omega_c = \omega - i \Gamma/2$ is a complex frequency~\cite{Kapusta:2006pm}. The imaginary part of this equation yields
\begin{equation}
    \Im \Pi = - \omega \Gamma .
\end{equation}
To interpret $\Gamma$, recall that a field evolves as $\Phi \sim e^{-i \omega_c t} \sim e^{-i \omega t} e^{- \Gamma t/2}$. The phase space distribution therefore evolves with time as $f_\Phi \sim \abs{\Phi}^2 \sim e^{-\Gamma t}$, with the condition that as $t\to \infty$, $f_\Phi \to f_\Phi^\mathrm{eq}$ where $f_\Phi^\mathrm{eq}$ is either the Bose-Einstein or the Fermi-Dirac distribution depending on the particle's spin. This means that
\begin{equation}
    f_\Phi(\omega, t) = f_\Phi^\mathrm{eq}(\omega) + c(\omega) e^{-\Gamma t}\,,
    \label{eq:transient}
\end{equation}
where $c(\omega)$ is a generic function of the energy. We see that $\Gamma(\omega) = -\Im\Pi(\omega)/\omega$ is the rate for $\Phi$ to equilibrate in the medium.

Now, consider the same distribution $f_\Phi$ evolving through the Boltzmann equation. Denoting the rates at which particles are absorbed and produced as $\Gamma_\mathrm{abs}$ and $\Gamma_\mathrm{prod}$ respectively, we obtain,
\begin{align}
\label{eq:rate}
    \frac{\mathrm{d}f_\Phi}{\mathrm{d}t} = -f_\Phi \Gamma_\mathrm{abs} + (1+\sigma f_\Phi)\Gamma_\mathrm{prod}\,,
\end{align}
where $\sigma\pm1$ accounts for Bose enhancement or Fermi suppression depending on the spin statistics of $\Phi$. If $\Gamma_\mathrm{abs}$ and $\Gamma_\mathrm{prod}$ do not depend on $f_\Phi$, this equation can be exactly solved to obtain,
\begin{align}
f_\Phi(\omega,\,t) = \frac{\Gamma_\mathrm{prod}}{\Gamma_\mathrm{abs} - \sigma \Gamma_\mathrm{prod}} + c(\omega) e^{-(\Gamma_\mathrm{abs}-\sigma\Gamma_\mathrm{prod}) t}\,.
\end{align}
If $\Phi$ is in thermal equilibrium with the heat bath, its phase space distribution is constant in time, $df_\Phi/dt = 0$. 
The principle of detailed balance can be used to derive a general relationship between $\Gamma_\mathrm{abs}$ and $\Gamma_\mathrm{prod}$ which are functions of the energy $\omega$ but not explicit functions of time,
\begin{align}
\label{eq:Gammaeq}
    \frac{\Gamma_\mathrm{prod}}{\Gamma_\mathrm{abs}} = \frac{f_\Phi^\mathrm{eq}}{1+\sigma f_\Phi^\mathrm{eq}} = e^{-\omega/T},\quad \quad \mathrm{with}\quad f_\Phi^\mathrm{eq} = \frac{1}{e^{\omega/T} -\sigma }\,.
\end{align}
From this we can write
\begin{align}
    f_\Phi(\omega, t) = f_\Phi^\mathrm{eq} + c(\omega) e^{-(\Gamma_\mathrm{abs}-\sigma\Gamma_\mathrm{prod}) t}\,,
\end{align}
from which we can immediately use Eq.~\eqref{eq:transient} to identify $\Gamma = \Gamma_\mathrm{abs}-\sigma\Gamma_\mathrm{prod}$. This links the imaginary part of the self-energy of $\Phi$ to its production rate,
\begin{align}
\label{eq:ImPiG}
    \mathrm{Im}\,\Pi = -\omega \Gamma = -\omega (\Gamma_\mathrm{abs}-\sigma\Gamma_\mathrm{prod}) = -\frac{\omega}{f_\Phi^\mathrm{eq}} \Gamma_\mathrm{prod}\,,
\end{align}
Using Eq.~\eqref{eq:ImPiG}, we can rewrite Eq.~\eqref{eq:rate} as,
\begin{align}
\label{eq:rateImPi}
    \frac{\mathrm{d}f_\Phi}{\mathrm{d}t} = -\frac{\mathrm{Im}\,\Pi}{\omega}(f_\Phi^\mathrm{eq}-f_\Phi).
\end{align}

Note that in the above discussion we have not made any assumptions as to whether or not $\Phi$ itself (or any other particle) is in equilibrium with the heat bath. However, while $\Phi$ can remain out of equilibrium for situations of interest, below we compute self-energies in the framework of the imaginary-time formalism (ITF), which is built on the assumption that all particles in the heat bath with which $\Phi$ interacts are in equilibrium. In diagrammatic terms, this means that all particles running in the loop of the $\Phi$ self-energy diagram must be in equilibrium. Therefore, the ITF can be used in the context of freeze-in or any other out-of-equilibrium production mechanism (i.e. with $f_\Phi \sim 0$) as long as $\Phi$ does not appear in its own self-energy loop. As discussed below in how we relate the self-energy to scattering processes, this means that the application of the ITF to the self-energy cannot describe the scattering or production of more than one $\Phi$ from the bath if $\Phi$ is out of equilibrium. If $\Phi$ is in equilibrium (or very close to it, $\abs{f_\Phi^\mathrm{eq}-f_\Phi}\ll 1$), then this formalism holds for all types of processes. For more general freeze-in production (with two or more particles being produced), it is necessary to use the real-time formalism \cite{Das:1997gg, Bellac:2011kqa}.

\subsection{Imaginary time formalism}
\label{sec:ITF}
For a background medium that is in equilibrium, observables at FTD can be evaluated using the ITF \cite{Das:1997gg, Bellac:2011kqa}. This formalism\footnote{For a pedagogical introduction to this topic in the context of BSM physics, including a detailed derivation of some of the results quoted in this work, see Ref.~\cite{Scherer:2024cff}.} is based on noting that the density matrix describing a system at temperature $T=1/\beta$, $n(\beta) = e^{-\beta \mathcal{H}}$, has the same mathematical form as the time evolution operator of Quantum Field Theory (QFT), $U(t,0) = e^{-it \mathcal{H}}$, where $\mathcal{H}$ is the time-independent Hamiltonian of the system. Because of this mathematical coincidence, any equilibrium observable at FTD can be calculated by using the machinery of vacuum QFT but with ``imaginary time,'' by
\begin{itemize}
    \item retaining the vacuum values and structure for all interaction vertices,
    \item replacing all energies by the corresponding discrete and imaginary bosonic ($\omega_n = i2n\pi T$) or fermionic ($\omega_n =i(2n+1)\pi T$) Matsubara frequencies, and,
    \item replacing integrals over energy with an infinite sum,
    \begin{align}
        \int \frac{d^4p}{(2\pi)^4}f(p_0, \vec{p}) \to T \sum_{n=-\infty}^{\infty}\int \frac{d^3p}{(2\pi)^3} f(\omega_n + \mu, \vec{p})\,,
    \end{align}
where $\mu$ is the chemical potential that encodes the effects of finite density in the medium. 
\end{itemize}  

\subsection{One-loop Example}
As an example, consider a generic scalar particle, $\Phi$, in a background of two other scalars, $\phi_1$ and $\phi_2$. The $\Phi$ self-energy can be written using the above ITF rules as,
\begin{equation}
\begin{aligned}
\adjincludegraphics[width=4cm,valign=c]{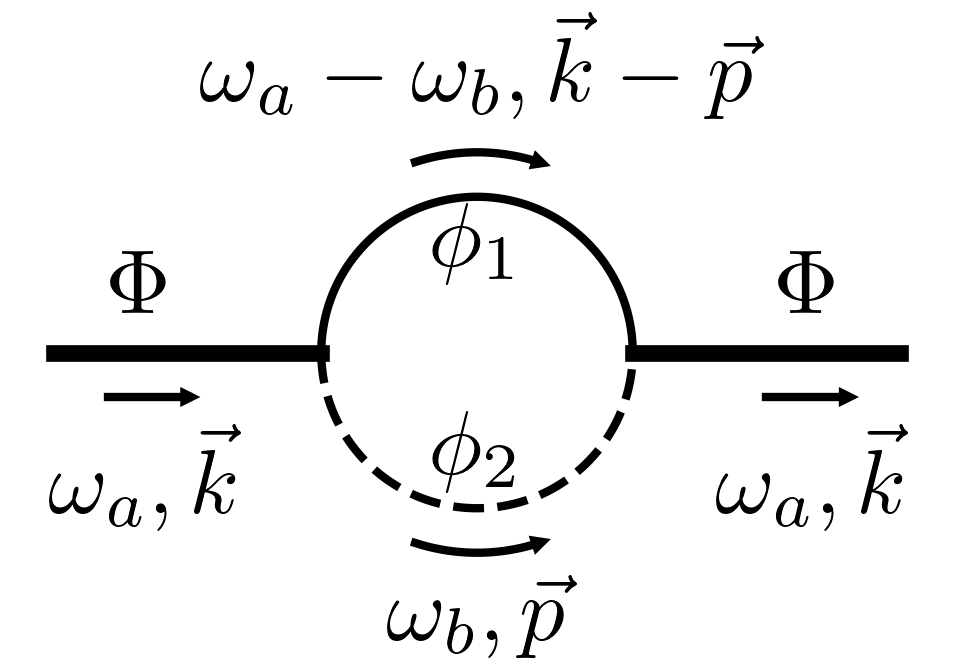} \equiv \Pi(\omega_a, \vec{k}) = g^2 T \sum_b \int\frac{d^3p}{(2\pi)^3} iD(\omega_b,\vec{p})iD(\omega_b-\omega_a, \vec{p}-\vec{k})\,,
\end{aligned}
\end{equation}
where, $\omega_b = 2i \pi b T$ is the bosonic Matsubara frequency and $iD(\omega_a, \vec{p}) = i/(\omega_a^2 - \vec{p}^2 - m^2)$ is the usual scalar propagator. On performing the infinite sum, we obtain \cite{Weldon:1983jn},
\begin{equation}
    \Pi(\omega_a, \vec{k}) \!= \!g^2 \!\int\!\!\frac{d^3p}{(2\pi)^3}\frac{1}{2E_1 2E_2}\frac{1+f_1 + f_2}{\omega_a - E_1 - E_2} + \frac{f_1-f_2}{\omega_a + E_1 - E_2} +\frac{f_2-f_1}{\omega_a - E_1  + E_2} - \frac{1+f_1+f_2}{\omega_a + E_1 + E_2},
\end{equation}
 where $f_{i}$ are the Bose-Einstein distribution functions, $f_i = (\mathrm{e}^{E_i/T}-1)^{-1}$ for particles $\phi_i$, and $E_i = \sqrt{\vec{p}_i^2 + m_i^2}$ are their corresponding energies. By extending this function to the full complex-$\omega$ plane, one can show that the corresponding expression has cuts along the real axis with discontinuities that are purely imaginary, $\mathrm{Disc}\Pi(\omega) = \lim_{\eta\to0}[\Pi(\omega + i\eta) - \Pi(\omega - i \eta)]= 2 i \mathrm{Im}\Pi(\omega)$ \cite{Bellac:2011kqa}, with
 \begin{align}
    \mathrm{Im}\Pi(\omega) = -\pi g^2 \int \frac{d^3p}{(2\pi)^3}\frac{1}{2E_12E_2}\{&\delta(\omega-E_1-E_2)[(1+f_1)(1+f_2)-f_1f_2] \nonumber \\
     &+ \delta(\omega+E_1-E_2)[f_1(1+f_2)-f_2(1+f_1)] \nonumber \\
     &+\delta(\omega-E_1+E_2)[f_2(1+f_1)-f_1(1+f_2)] \nonumber \\
     &+\delta(\omega+E_1+E_2)[f_1f_2-(1+f_2)(1+f_1)]
    \} \,,
\end{align}
where factors of $f_1f_2$ have been added and subtracted to yield the phase space factors above. By strategically relabeling $p \to p_i$ with $i=1$ or $2$ and inserting fat unity $\mathbb{1} = \int d^3p_{i} \delta^{(3)}(\vec{k}\pm\vec{p_1}\pm\vec{p_2})$, it is straightforward to show how the imaginary part of the self-energy is related to tree-level production and annihilation processes, 
\begin{align}
\mathrm{Im}\Pi(\omega, \vec{k}) = -\frac{1}{2} \int &\frac{d^3p_1}{(2\pi)^32E_1} \frac{d^3p_2}{(2\pi)^32E_2}\\
   &\times\bigg\{(2\pi)^4 \delta^{(4)}(k - p_1 -p_2)|M|^2_{\Phi \to \phi_1\phi_2}[(1+f_1)(1+f_2)-f_1f_2]\nonumber \\
   &+(2\pi)^4 \delta^{(4)}(k+p_1-p_2)|M|^2_{\Phi\phi_1 \to \phi_2}[f_1(1+f_2)-f_2(1+f_1)] \nonumber \\
   &+(2\pi)^4\delta^{(4)}(k-p_1+p_2)|M|^2_{\Phi\phi_2\to\phi_1}[f_2(1+f_1)-f_1(1+f_2)] \nonumber \\
    &+(2\pi)^4\delta^{(4)}(k+p_1+p_2)|M|^2_{\Phi\phi_1\phi_2 \to 0}[f_1f_2-(1+f_2)(1+f_1)]
   \bigg\}\, .\nonumber\label{eq:ImPiScalar1loop}
\end{align}
This explicitly reproduces Eq.~\eqref{eq:ImPiG} at the one-loop level. The interaction rates for a particle in a medium can therefore be obtained diagrammatically from a particle's self-energy by ``cutting'' the diagram, or equivalently putting all the particles running in the loop on-shell, and integrating over the relevant thermal phase space,\footnote{Note that in the diagrammatic equations in this work, such as Eq.~\eqref{eq:ImPiScalar1loop_diag}, the thermal phase space integrals for the processes on the right hand side have been dropped for brevity. These can be inferred from the corresponding equations provided in the text.} 
\begin{equation}\label{eq:ImPiScalar1loop_diag}
\begin{aligned}
   & \adjincludegraphics[height = 2cm, valign=c]{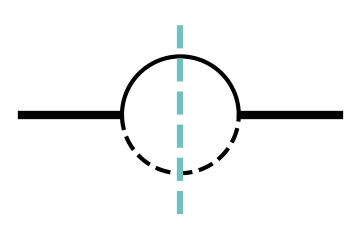}
    = \qty[\adjincludegraphics[height=1.5cm, valign=c]{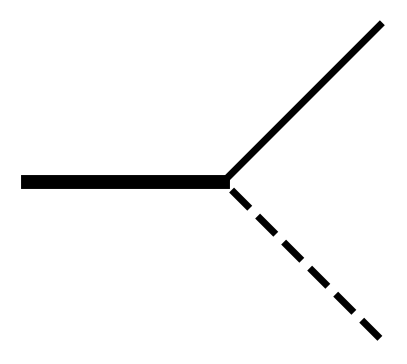}
    \times \adjincludegraphics[height=1.5cm, valign=c]{diagrams/Scalar-acut-0left.png}^{*}]
    + \qty[\adjincludegraphics[height=1.5cm, valign=c]{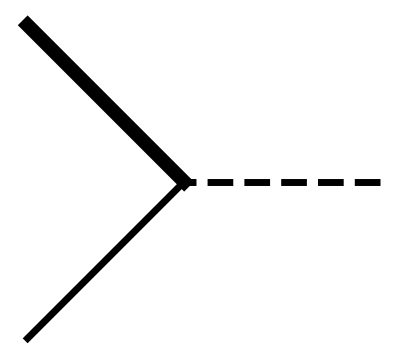}
    \times \adjincludegraphics[height=1.5cm, valign=c]{diagrams/Scalar-acut-1left.png}^{*}]\\
    &\hspace{3.2cm}+ \qty[\adjincludegraphics[height=1.5cm, valign=c]{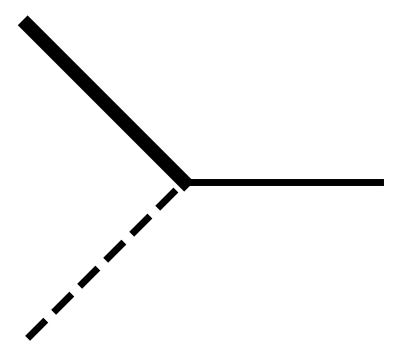}
    \times \adjincludegraphics[height=1.5cm, valign=c]{diagrams/Scalar-acut-2left.png}^{*}]
    + \qty[\adjincludegraphics[height=1.5cm, valign=c]{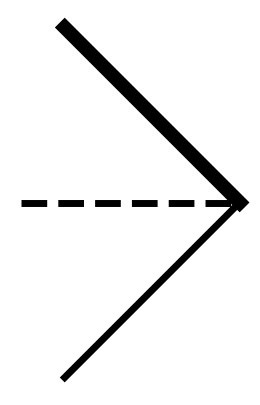}
    \times \adjincludegraphics[height=1.5cm, valign=c]{diagrams/Scalar-acut-3left.png}^{*}] - (p_i \to -p_i)
    \\
    &\hspace{0.5cm}=\abs{\adjincludegraphics[height=1.5cm, valign=c]{diagrams/Scalar-acut-0left.png}}^2
    +\abs{\adjincludegraphics[height=1.5cm, valign=c]{diagrams/Scalar-acut-1left.png}}^2
    +\abs{\adjincludegraphics[height=1.5cm, valign=c]{diagrams/Scalar-acut-2left.png}}^2
    +\abs{\adjincludegraphics[height=1.5cm, valign=c]{diagrams/Scalar-acut-3left.png}}^2 - (p_i \to -p_i), \qquad 
\end{aligned}
\end{equation}
where $^*$ denotes the complex conjugate of the diagram. This result is analogous to the optical theorem in vacuum QFT, where the Cutkosky rules relate the imaginary part of the self-energy to an interaction cross-section \cite{cutkosky1960singularities, Peskin:1995ev, Schwartz:2014sze}. However, contrary to the optical theorem in a vacuum, the particles that are put on shell by the cut can act as initial state particles and therefore we get \textit{both} production and absorption in a medium. This is encapsulated in the $(p\to-p)$ term above. Further, not all of the processes drawn in Eq.~\eqref{eq:ImPiScalar1loop_diag} are kinematically allowed. For example, $\Phi \phi_1\phi_2 \leftrightarrow 0$ is prohibited by energy conservation, and therefore will not contribute to the total damping rate.

A similar calculation can be performed for fermion self-energies. The discontinuity in this case is given by $\mathrm{Im}\Pi(\omega, \vec{k}) = - \omega(\Gamma_\mathrm{abs} + \Gamma_\mathrm{prod})$ \cite{Weldon:1983jn}. Consequently, for any particle, production via decays or inverse decays in a medium at leading order (corresponding to cuts through one-loop self-energy diagrams) can be exactly evaluated by considering the zero-temperature tree-level amplitudes permitted by the theory, appropriately weighted by thermal phase space factors. 

\subsection{Generalizing to multi-loop diagrams}
A common assumption in the BSM phenomenology literature is that this correspondence between vacuum and FTD rates holds for loops of higher order (for instance, $2\to2$ processes would be the result of cutting a two-loop self-energy diagram) and therefore, particle production in a medium can be understood \textit{purely} in terms of the processes that happen in a vacuum with an appropriately modified phase space. As an example of applying this correspondence, dark photons in an electron-photon plasma would have a total interaction rate given by a sum of the net decay, Compton, and fusion rates~\cite{Hardy:2016kme}, 
\begin{equation}
    \begin{aligned}
        &\adjincludegraphics[height=1.95cm,valign=c]{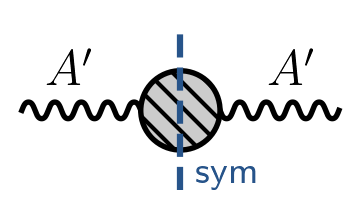} 
        = \adjincludegraphics[height=1.95cm,valign=c]{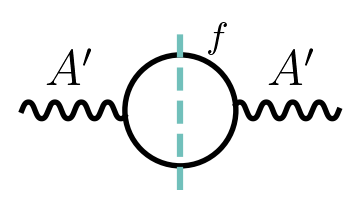}
        + \adjincludegraphics[height=1.95cm,valign=c]{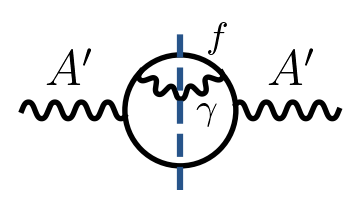}
        + \adjincludegraphics[height=1.95cm,valign=c]{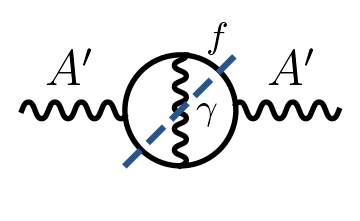}\\
        &\qquad= 
        - \abs{\adjincludegraphics[height=1.95cm,valign=c]{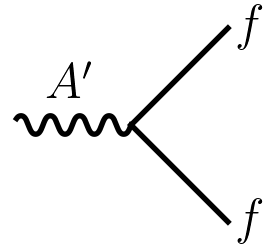}}^2
        - \abs{\adjincludegraphics[height=1.95cm,valign=c]{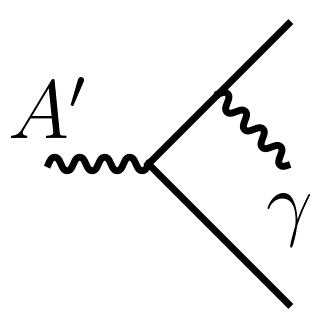}}^2
        - \abs{\adjincludegraphics[height=1.95cm,valign=c]{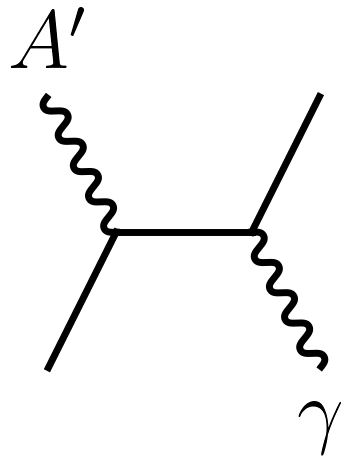} 
            + \adjincludegraphics[height=1.95cm,valign=c]{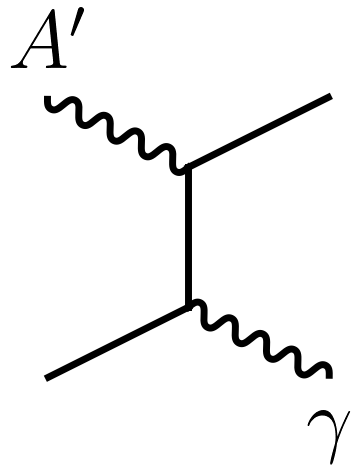}}^2
        - \abs{\adjincludegraphics[height=1.95cm,valign=c]{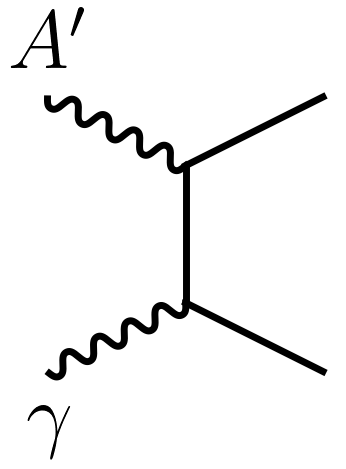}}^2\\
        &\qquad \quad+ \abs{\adjincludegraphics[height=1.95cm,valign=c]{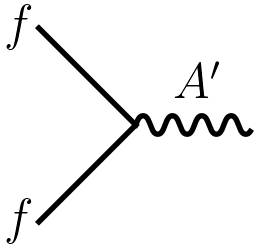}}^2
        + \abs{\adjincludegraphics[height=1.95cm,valign=c]{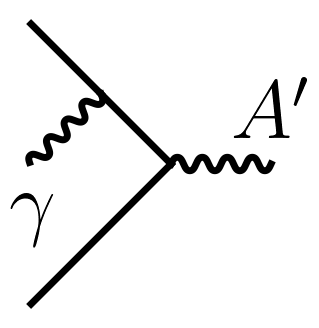}}^2
        + \abs{\adjincludegraphics[height=1.95cm,valign=c]{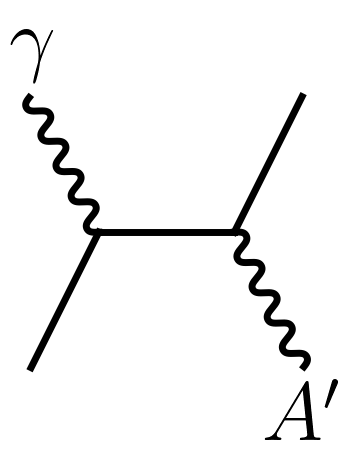} 
            + \adjincludegraphics[height=1.95cm,valign=c]{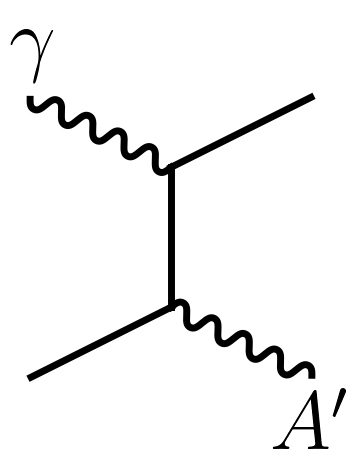}}^2
        + \abs{\adjincludegraphics[height=1.95cm,valign=c]{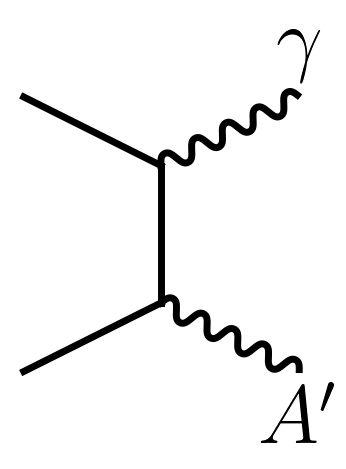}}^2
        .
    \end{aligned}
    \label{eq:Aprimecuts}
\end{equation}
Much of the literature on BSM particle production in plasmas relies on this assumption. However, as we show in this work, this assumption breaks down for $n>1$ loops (or correspondingly, production channels involving more particles than just a simple decay). Diagrammatically, this is because the cuts presented in Eq.~\eqref{eq:Aprimecuts} do not completely account for all possible physical cuts. In particular, for multi-loop diagrams, there can be asymmetrical cuts, 
\begin{equation}
    \begin{aligned}
        \adjincludegraphics[height=1.95cm,valign=c]{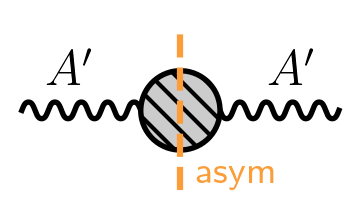} 
         &=  
        \adjincludegraphics[height=1.95cm,valign=c]{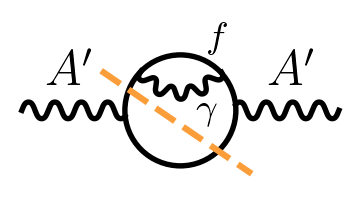}
        + \adjincludegraphics[height=1.95cm,valign=c]{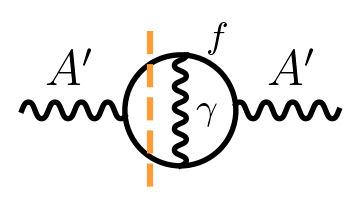},
    \end{aligned}
    \label{eq:Aprime_asym1}
\end{equation}
which correspond to discontinuities that arise as a result of an \textit{interference} between a usual vacuum amplitude (left half of the cut diagrams in Eq.~\eqref{eq:Aprime_asym1}) and one that has to be corrected because of the presence of background fields (right half of the cut diagrams). Such processes clearly do not contribute to the scattering rate in a vacuum: the cut from the first term on the RHS in Eq.~\eqref{eq:Aprime_asym1} yields an amplitude with a loop on an external leg which has to be amputated when calculating scattering processes~\cite{Peskin:1995ev}, and the cut arising from the third term is a vertex correction accounted for by renormalization. In a medium however, the loops in these cut diagrams correspond to physical exchanges of real particles present in the medium and as a result cannot be amputated or renormalized in the usual fashion. Instead, the corresponding amplitudes contribute non-trivially to the scattering rate. Conceptually, these interference-type processes can be understood as a form of \textit{forward scattering} on the background medium. For example, the first cut in Eq.~\eqref{eq:Aprime_asym1} can be written as the product of a decay amplitude with a spectator photon (or fermion) field from the background medium and a scattering amplitude where the photon (or fermion) from the medium has the same incoming and outgoing four-momentum and is denoted in yellow,
\begin{equation}
    \begin{aligned}
        \adjincludegraphics[height=1.95cm,valign=c]{diagrams/Aprime_asym_leg.png} 
        &\supset \left(\adjincludegraphics[height=2.5cm,valign=c]{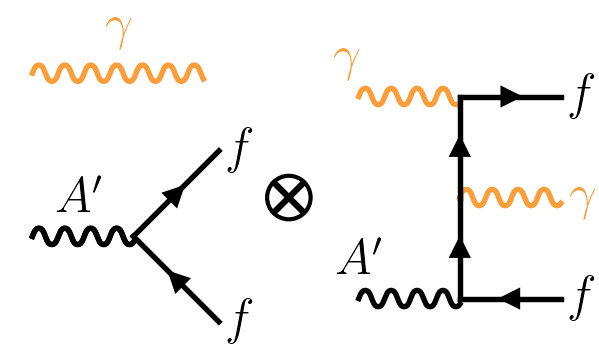}\right)
        + \left(\adjincludegraphics[height=2.5cm,valign=c]{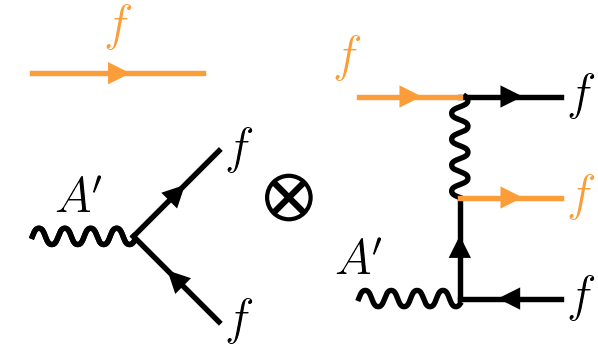}\right).
    \end{aligned}
    \label{eq:Aprime_interf}
\end{equation}
Note that this is different from how forward scattering is usually defined since \textit{only} the background spectator field maintains its momentum. Further, the spectator field is added by hand to the left halves of the cut diagrams in Eq.~\eqref{eq:Aprime_interf} as a visual aid to demonstrate the presence of the medium, but it does not enter in the Feynman amplitudes for these halves (see Sec.~\ref{sec:SSHBrules} for details).

The existence of these diagrams has been pointed out in the thermal field theory literature~\cite{Wong:2000hq, Majumder:2001iy}, but their impact on BSM phenomenology has mostly been overlooked and has not been quantified. In the following Section, we provide a complete set of rules to parameterize all production processes that contribute at finite temperature using the ITF. These rules are valid up to arbitrarily high loop order in the particle self-energy, and can be used to accurately model particle production in a medium. We further apply these rules to two toy models and quantify the effect of interference diagrams on BSM particle production.

\section{Rules for Spectral Surgery in a Heat Bath (SSHB)}
\label{sec:SSHBrules}
\subsection{Summary of SSHB}
The aim for SSHB is to reduce $n$-loop self-energy graphs to thermally corrected tree-level processes so that the impact of the medium on observables can be quantified without resorting to a full FTFT calculation. In this section, we present the cutting rules at FTD in a consolidated form, leaving their corroboration with thermal field theory for subsequent sections. These rules can be used to calculate the total damping rate of a particle $\Phi$ in a medium in the following way:
\begin{itemize}
    \item \textbf{Step 0. Self-energy diagrams:} Draw all the Feynman diagrams contributing to the self-energy of $\Phi$ at the required loop order.
    \item \textbf{Step 1. Bisection:} For each diagram, list all distinct bisections, or all distinct cuts, both symmetric and asymmetric, that separate the self energy into two parts, with each part containing one copy of $\Phi$.
    \item \textbf{Step 2. Converting bisections into processes:} For each bisection, list all the different processes that arise by considering the cut propagators as particles either in the initial or final state. 
    \item \textbf{Step 3. Loopectomy:} For each of these processes, open up all internal loops, if any, using partial cuts to obtain tree-level amplitudes; the cut propagators in this case become on-shell \textit{spectator} particles. List all possible partially cut diagrams from loopectomy, i.e. by enumerating all the different loop openings and all permutations of initial and final states for spectators.
    \item \textbf{Step 4. Sum over tree-level processes:} For each bisection, evaluate the contribution from each process and sum over all processes using Eq.~\eqref{eq:PiN_cut} and Eq.~\eqref{eq:amp_full}. Finally sum over all bisections. 
\end{itemize}
At the end of this Section, we provide a brief discussion of various subtleties that arise in this framework.

\subsection{Worked example: scalar toy model}
\begin{table}
\begin{center}
\begin{tabular}{ |c|c| } 
 \hline
 Variable  & Definition \\ 
 \hline
 $n$  & Indexes loop order under consideration \\ 
 \hline
 $d$  & Indexes self-energy diagrams at a given loop order \\ 
 \hline
 $b$ & Indexes bisections of a self-energy diagram \\
 \hline
 $l_b$ & Indexes number of loop propagators cut for a bisection \\
 \hline
 $c$ & Indexes processes arising out of a bisection \\
 \hline
  $\tilde c$ & Indexes processes arising out of partial cuts \\
  \hline
 $s_i^{c,\tilde{c}}$ & Denotes if particle $i$ of a process is incoming or outgoing\\ 
 \hline
$F_{c}$ & Thermal phase space for interacting particles \\
\hline
$\mathcal{F}_{\tilde c}$ & Thermal phase space for spectator particles \\ 
\hline
$\mathcal{S}_{c,\tilde{c}}$ & Set of all external state particles excluding the test particle $\Phi$ for a process \\
\hline
\end{tabular}
\caption{Definitions of some key variables used in the worked example. \label{tab:def}}
\end{center}
\end{table}

To provide an explicit example of how to use these rules, we consider a toy model of scalar particles with the interaction Lagrangian
\begin{equation}
    \mathcal{L}_\text{int} = g \Phi \phi_1 \phi_2 + \frac{1}{2} \lambda \phi_1^2 \phi_3\,,
    \label{eq:scalar_lag}
\end{equation}
where $\Phi$ is the particle whose self-energy is to be evaluated, and $\phi_1,\,\phi_2,\,\phi_3$ are background particles assumed to be in thermal equilibrium. We assume $\phi_2$ is kept in equilibrium through some other process maybe not captured by this Lagrangian. This Lagrangian results in a single two-loop self-energy diagram, which we will consider as a working example throughout the rest of this section. The one-loop self-energy for this model was already calculated in Ref.~\cite{Weldon:1983jn} and is additionally presented in Section \ref{sec:ITF}. For convenience, we provide a list of the definitions of the key variables used in this section in Table~\ref{tab:def}.

\begin{step}\label{step0}\end{step}\textbf{Self-energy diagrams: } 
We denote the $n$-loop self-energy for a particle $\Phi$ with four-momentum $k = (\omega, \vec{k})$ in a thermal bath as $\Pi^{(n)}(k)$. In general, this consists of a sum of multiple self-energy diagrams indexed by $d$. The imaginary part of the $n$-loop self-energy can thus be expressed as 
\begin{equation}
    \Im \Pi^{(n)}(k) = \sum_{d \in \text{diagrams}} \Im \Pi^{(n,d)}(k)
\end{equation}
The model we are considering results in a single 2-loop self-energy diagram:
\begin{equation}
\adjincludegraphics[height=2.4cm, valign=c]{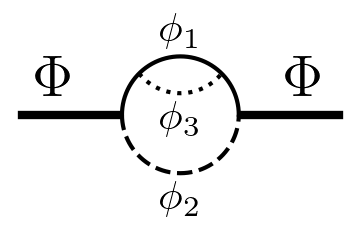}
\label{eq:2-loop-scalar}
\end{equation}
In contrast, the model in Eq.~\eqref{eq:Aprimecuts} would, for example, have three two-loop self-energy contributions, which would need to be summed over.

\begin{step}\label{step1}\end{step}\textbf{Bisection: }
The imaginary part of $\Pi^{(n,d)}(k)$ arises from \textit{all distinct ways} of splitting the self-energy diagrams in two parts, such that each component of the diagram contains a $\Phi$ state. We refer to these types of splits as \textit{bisections}, implying that 
\begin{equation}
    \Im\Pi^{(n,d)}(k) = \sum_{b \in \text{bisection}} \Im\Pi^{(n,d)}_b(k).
    \label{eq:possible_cuts}
\end{equation}
For our 2-loop diagram, the bisections are
\begin{align}
    \Im &\left(\adjincludegraphics[height=2.0cm, valign=c]{diagrams/Scalar-2loop-none.png}\right) = \mathop{\adjincludegraphics[height=2.0cm, valign=c]{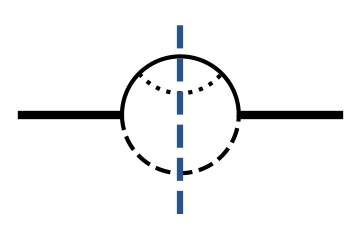}}_{\mathrm{I}} + \mathop{\adjincludegraphics[height=2.0cm, valign=c]{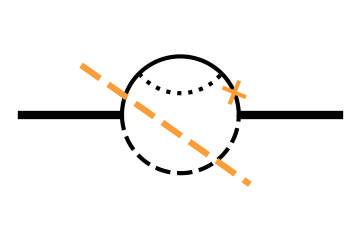}}_{\mathrm{II}}
    \label{eq:scalar2loop_bisection}
\end{align}
where the colored, dashed lines denote the cuts through one or more propagators. Cutting a propagator means putting the corresponding particle on-shell. Hence, if two or more propagators in a diagram share the same momentum and correspond to identical particles, \textit{then putting one on-shell automatically puts the others on-shell as well}. This has two consequences. First, one does not need to distinguish between bisections which put the same propagators on-shell, since they correspond to the same physical process. In our two-loop example, there are thus only two distinct bisections I and II, as the following two ways of splitting the diagram are equivalent:
\begin{align}
    \adjincludegraphics[height=2.3cm, valign=c]{diagrams/Scalar-2loop-asym.png} \equiv \adjincludegraphics[height=2.3cm, valign=c]{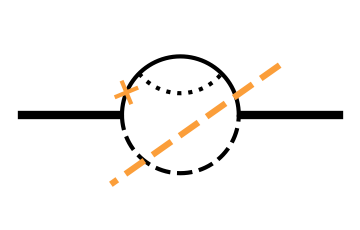}
    \label{eq:2cut_equiv}
\end{align}
Second, cutting a diagram may result in amplitudes which have \textit{internal states} that are on-shell. In our example, these are denoted by the orange x's in Eqs.~\eqref{eq:scalar2loop_bisection} and ~\eqref{eq:2cut_equiv}. 

Summarizing step 1, the sum over $b$ in Eq.~\eqref{eq:possible_cuts} runs over all distinct (non-redundant) sets of bisections, ensuring that each physical process is counted only once, and may include amplitudes with on-shell internal states. We now turn to evaluating the contribution to $\Im\Pi$ from each distinct bisection. 

\begin{step}\label{step2}\end{step}\textbf{Converting bisections into processes: }
For a given bisection $b$, each propagator that is put on-shell can become an \textit{incoming} or \textit{outgoing} external state. For the purposes of these rules, we treat the external $\Phi$ particle as \textit{incoming} so that we consider only absorption amplitudes diagrammatically. Assuming CP symmetry, the inverse processes are accounted for by simply modifying the thermal phase space (see the discussion around Eq.~\eqref{eq:Fac} for more details). As a result, a single bisection $b$ cutting through $l_b$ loop propagators gives rise to a sum of $ 2^{l_b}$ \textit{processes} corresponding to all possible permutations of incoming and outgoing states. Therefore, a process, $c$, is defined both by the choice of propagators put on-shell (i.e. the choice of bisection) \textit{and} by the choice of which particles are incoming and outgoing.
In our 2-loop example, the first bisection $b=\mathrm{I}$ (term (I) in Eq.~\eqref{eq:scalar2loop_bisection}) cutting through $l_b=3$ propagators corresponds to the following $2^{3}=8$ processes: 
\begin{equation}
\resizebox{0.75\textwidth}{!}{
$\displaystyle
\begin{aligned}
\mathop{\adjincludegraphics[height=2.3cm, valign=c]{diagrams/Scalar-2loop-sym.png}}
    =&\mathop{\abs{\adjincludegraphics[width=2.1cm, valign=c]{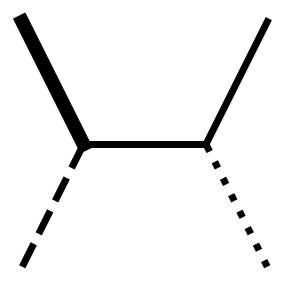}}^2}
    +\mathop{\abs{\adjincludegraphics[width=2.1cm, valign=c]{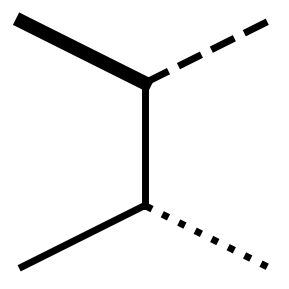}}^2}
    +\mathop{\abs{\adjincludegraphics[width=2.1cm, valign=c]{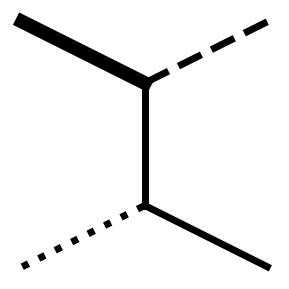}}^2}
    \\
    +&\mathop{\abs{\adjincludegraphics[width=2.1cm, valign=c]{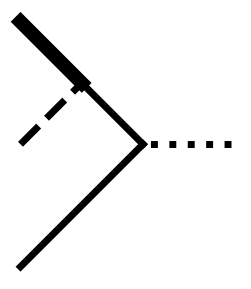}}^2}
    +\mathop{\abs{\adjincludegraphics[width=2.1cm, valign=c]{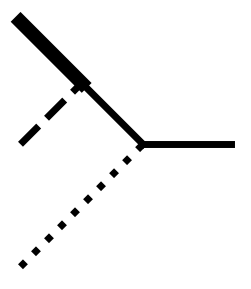}}^2}
    +\mathop{\abs{\adjincludegraphics[width=2.1cm, valign=c]{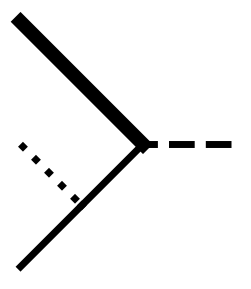}}^2}
    \\
    +&\mathop{\abs{\adjincludegraphics[width=2.1cm, valign=c]{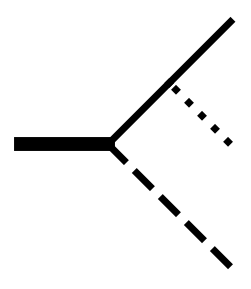}}^2}
    +\mathop{\abs{\adjincludegraphics[width=2.1cm, valign=c]{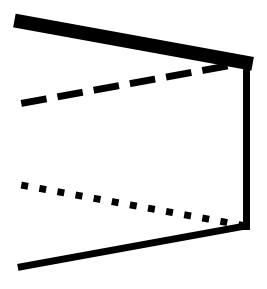}}^2}
\end{aligned}
$}
\label{eq:sym_cut_2loop}
\end{equation}
while the second bisection $b=\mathrm{II}$ (term (II) in Eq.~\eqref{eq:scalar2loop_bisection}) cutting through $l_b=2$ propagators corresponds to the following $2^{2}=4$ processes:
\begin{equation}
\begin{aligned}
    \mathop{\adjincludegraphics[height=2.2cm, valign=c]{diagrams/Scalar-2loop-asym.png}}
    = &\mathop{\adjincludegraphics[width=4cm, valign=c]{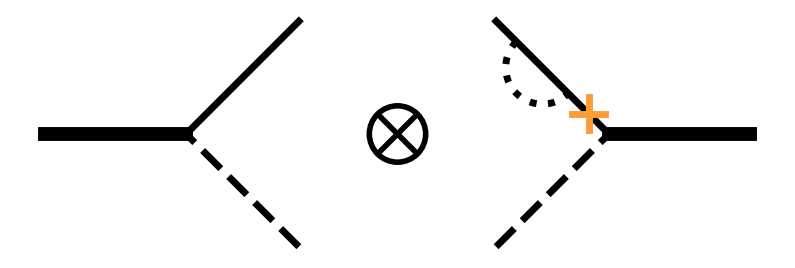}}
    + \mathop{\adjincludegraphics[width=4cm, valign=c]{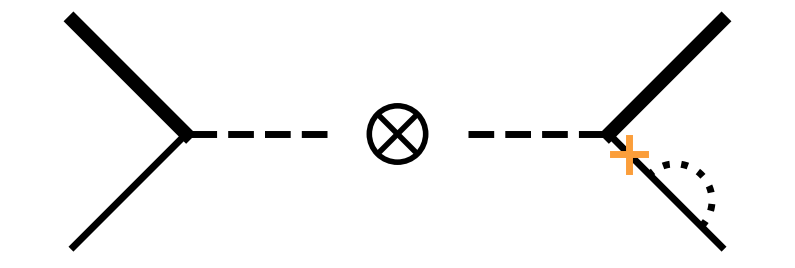}}
    \\
    + &\mathop{\adjincludegraphics[width=4cm, valign=c]{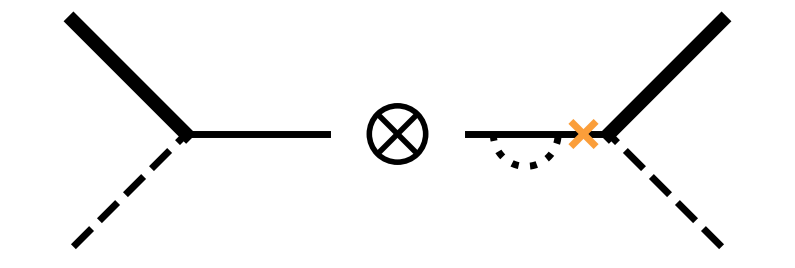}}
    + \mathop{\adjincludegraphics[width=4cm, valign=c]{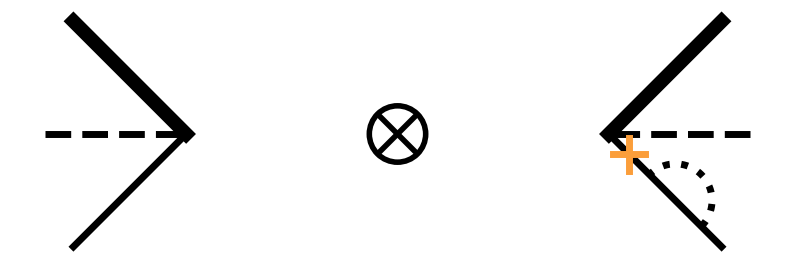}}.
\end{aligned}
\label{eq:internal_loop1}
\end{equation}
We note that the right halves of these bisected diagrams are equivalent to the complex conjugate of the amplitude where all incoming and outgoing states are interchanged, $\mathcal{M}_{n \to m} = \mathcal{M}_{m\to n}^*$.
In general, the contribution from a specific bisection $b$ can be written as a sum of these $2^{l_b}$ processes,
\begin{equation}
    \Im\Pi^{(n,d)}_b(k)
    = -\frac{1}{2} \sum_{c=1}^{2^{l_b}}
    \int d\Phi_{\alphac}
    \deltabar^{4}\left(k+\sum_{i}s^{\alphac}_{i}p_{i}\right)
    F_{\alphac}
    \sum_{\mathrm{spins}}\left[\tilde{\M}_{L}^{\alphac} \tilde{\M}_R^{\alphac}\right]\,,
\label{eq:PiN_cut}
\end{equation}
with the following constituents:
\begin{itemize}
    \item \underline{Phase space measure}:
The phase space measure is given by 
\begin{equation}
d\Phi_{\alphac}\equiv \prod_{i\in \mathcal{S}_\alphac}\dkbar[4]{p_{i}}\frac{(-1)^{n_{i}-1}}{(n_{i} -1)!}\deltabar^{(+)(n_{i}-1)}\left(p_{i}^{2}-m_{i}^{2}\right)\,.
\end{equation}
In the equation above, $\mathcal{S}_\alphac$ is the set of \textit{all} external state particles \textit{except} the test particle $\Phi$, $n_i$ is the number of identical propagators for particle of type $i \in \mathcal{S}_\alphac$ that have been put on-shell and $\dkbar[4]{p} \equiv d^4 p/(2\pi)^4$.
As for the $\delta$-function,  $\deltabar(\ldots) \equiv 2\pi\delta(\ldots)$, the $(+)$ indicates we pick up the positive energy solution, and the $(n_i-1)$ superscript indicates we are taking the $(n_i-1)^\mathrm{th}$ derivative with respect to $p_0^2$, i.e.
\begin{align}
    \deltabar^{(+)(n_i-1)}\left(p_{i}^{2}-m_{i}^{2}\right) \equiv (2\pi) \theta(p_i^0)\delta^{(n_i-1)}\left(p_{i}^{2}-m_{i}^{2}\right)
\end{align}
The $(n_i-1)^\text{th}$ derivative arises because a cut can put \textit{internal particles} on shell as well, as discussed in Step~\ref{step1}. For example, each term in Eq.~\eqref{eq:internal_loop1} would include a factor of $\delta^{(+)(1)}(p_{\phi_1}^{2}-m_{\phi_1}^{2})\equiv \delta^{(+)'}(p_{\phi_1}^{2}-m_{\phi_1}^{2})$ in their phase space, since cutting through one of the $\phi_1$'s in the loop puts the other one on-shell, i.e., $n_i=2$. It is straightforward to see that if the cut passes through all unique propagators as in Eq.~\eqref{eq:sym_cut_2loop}, or equivalently if $n_i=1$, we recover the expected result similar to the optical theorem of vacuum QFT.

\item\underline{Momentum conserving delta function}: The delta function, $\deltabar^{4}\left(k+\sum_{i}s^{\alphac}_{i}p_{i}\right)$ is determined by the choice of incoming and outgoing states and therefore specified by the process $\alphac$. Here, $s^{\alphac}_i = +1,-1$ denotes that the particle $i$ with momentum $p_i$ is incoming or outgoing, respectively (assuming that $\Phi$ has four-momentum $k$). 
\newpage

\item \underline{Particle distribution factor}:
The weights from particle distribution functions in the thermal bath for each process are encapsulated by $F_{\alphac}$, which is given by
\begin{equation}
F_{\alphac}\equiv\prod_{i\in \mathcal{S}_\alphac} \left[\theta(s_{i}^c)f^{(\eta_{i})}(\abs{p_{i}^{0}})+\theta(-s_{i}^c)(1+\eta_{i} f^{(\eta_{i})}(\abs{p_{i}^{0}}))\right] + \eta_\Phi (s_i^c \rightarrow - s_i^c)
\label{eq:Fac}
\end{equation}
where $f^{(\eta_i)}(|p_i^0|) = (e^{|p_i^0|/T}+\eta_i)^{-1}$ denotes phase space distributions for bosons ($\eta_i=-1$) and fermions ($\eta_i=1$). We therefore get a factor of $f^{(\eta_i)}(|p_i^0|)$ 
for incoming  states and a factor of $\left(1+\eta_if^{(\eta_i)}(|p_i^0|)\right)$ for outgoing states. These factors arise naturally from the finite temperature field theory treatment, as shown in the next section and correspond to the usual thermal phase space for interacting particles \cite{Gondolo:1990dk}.
The first term in Eq.~\eqref{eq:Fac} corresponds to $\Phi$ \textit{absorption}. If the matrix element squared of a process is assumed to be CP symmetric, $|\mathcal{M}|^2_{n\to m} = |\mathcal{M}|^2_{m\to n}$, production rates
can be easily incorporated by switching $s_i^c\to-s_i^c$ in the particle distribution factor. To obtain the net production rate, the two terms have to be added if $\Phi$ is a fermion ($\eta_\Phi=1$) and subtracted if $\Phi$ is a boson  ($\eta_\Phi=-1$). This gives us the second term of Eq.~\eqref{eq:Fac} (see also the discussion in Ref.~\cite{Weldon:1983jn}).

\item \underline{Matrix element}: Each process $\alphac$ is a product of two amplitudes $\tilde{\M}_{L}^\alphac$ and $\tilde{\M}_{R}^\alphac$ corresponding to the left and right halves of a bisected diagram (see for instance Eq.~\eqref{eq:internal_loop1}). We define these as
\begin{equation}
    \tilde{\M}_{L}^{\alphac} \tilde{\M}_R^{\alphac} = {\M}_{L}^{\alphac} {\M}_R^{\alphac} \times
    \prod_{i\in \mathcal{S}_\alphac}  
    (p_i^2 - m_i^2)^{n_i-1}
    \label{eq:tildeMatrixElements}
\end{equation}
where $\M_{L,R}^\alphac$ is the matrix element calculated by using vacuum Feynman rules for tree level diagrams and the methodology described in Step~\ref{step3}
for loop diagrams. The product $\prod_{i\in \mathcal{S}_\alphac}(p_i^2 - m_i^2)^{n_i-1}$ regulates the amplitude for \textit{on-shell} internal states, if any (denoted by the x's in our diagrams). 
\end{itemize}
If a bisection results in processes with purely tree-level amplitudes, such as in Eq.\eqref{eq:sym_cut_2loop}, we recover the well-known thermal interaction rate of Ref.~\cite{Weldon:1983jn} upon applying Eq.~\eqref{eq:PiN_cut},
\begin{equation}
\begin{aligned}
\mathop{\adjincludegraphics[height=1.7cm, valign=c]{diagrams/Scalar-2loop-sym.png}}=\\
    -\frac{ g^2 \lambda^2 }{2}
    \int \dkbar[4]{p_2}& \deltabar^{(+)}(({p_2^0})^2-E_{p_2}^2)
    \int \dkbar[4]{p_3} \deltabar^{(+)}(({p_3^0})^2-E_{p_3}^2)
    \int \dkbar[4]{p_1} \deltabar^{(+)}(({p_1^0})^2-E_{p_1}^2)\\
    \times\Big[ \ 
    &\deltabar^{4}(k + {p_2} - {p_3} - {p_1}) \qty[D(k + {p_2})]^2 \qty[f_{p_2} (1+f_{p_3}) (1+f_{p_1}) - f_{p_3} f_{p_1} (1+f_{p_2})]\\
    +&\deltabar^{4}(k - {p_2} - {p_3} + {p_1}) \qty[D(k - {p_2})]^2 \qty[f_{p_1} (1+f_{p_2}) (1+f_{p_3}) - f_{p_2} f_{p_3} (1+f_{p_1})]\\
    +&\deltabar^{4}(k - {p_2} + {p_3} - {p_1}) \qty[D(k - {p_2})]^2 \qty[f_{p_3} (1+f_{p_2}) (1+f_{p_1}) - f_{p_2} f_{p_1} (1+f_{p_3})]\\
    +&\deltabar^{4}(k + {p_2} - {p_3} + {p_1}) \qty[D(k + {p_2})]^2 \qty[f_{p_2} f_{p_1} (1+f_{p_3}) - f_{p_3} (1+f_{p_1}) (1+f_{p_2})]\\
    +&\deltabar^{4}(k + {p_2} + {p_3} - {p_1})\qty[D(k + {p_2})]^2 \qty[f_{p_2} f_{p_3} (1+f_{p_1}) - f_{p_1}(1+f_{p_2})(1+f_{p_3})] \quad\\
    +&\deltabar^{4}(k - {p_2} + {p_3} + {p_1}) \qty[D(k - {p_2})]^2 \qty[f_{p_3} f_{p_1} (1+f_{p_2}) - f_{p_2} (1+f_{p_3}) (1+f_{p_1})]\\
    +&\deltabar^{4}(k - {p_2} - {p_3} - {p_1}) \qty[D(k - {p_2})]^2 \qty[(1+f_{p_2}) (1+f_{p_3}) (1+f_{p_1}) - f_{p_2} f_{p_3} f_{p_1}]\\
    +&\deltabar^{4}(k + {p_2} + {p_3} + {p_1}) \qty[D(k + {p_2})]^2 \qty[f_{p_2} f_{p_3} f_{p_1} - (1+f_{p_2}) (1+f_{p_3}) (1+f_{p_1})]
    \ \Big]\\
\end{aligned}
\end{equation}
where the eight terms in this expression correspond to the eight processes shown diagrammatically in Eq.~\eqref{eq:sym_cut_2loop}.

However, bisections can result in processes where either or both $\tilde{\M}_{L}^\alphac$ and $\tilde{\M}_{R}^\alphac$ are multi-loop amplitudes (see for example, Eq.~\eqref{eq:internal_loop1}. We call 
these loops \textit{internal thermal loops} and discuss how to evaluate such amplitudes in the following step. 

\begin{step}\label{step3}\end{step}\textbf{Loopectomy (opening up internal loops): } 
In the following, we will suppress the $L,\,R,$ indices for the diagrams corresponding to the various processes in Step~\ref{step2}. For a given diagram, we denote the total number of internal loops with $n_l$.
We open these internal loops through \textit{partial cuts}, by putting $n_l$ 
propagators on-shell such that the full diagram 
remains connected. For each cut propagator, we must put one leg in the incoming state and one in the outgoing state. This is because only the real part of these internal loop diagrams contribute to the imaginary part of the self-energy.
These partially cut diagrams
correspond to the absorption and emission of a background particle with the \textit{same incoming and outgoing momentum}, analogous to scattering in the forward limit. We refer to the particles being put on-shell during this ``loopectomy'' as \textit{spectators}.  A partially cut diagram
$\tilde{c}$ is therefore defined by the choice of which $n_l$ propagators are put on-shell as spectators \emph{and} the choice of which spectator legs are in the incoming and outgoing states.
Through opening these loops, we can therefore reduce an $n_l$-loop diagram into a sum of tree-level partially cut diagrams, properly weighed by phase space factors. So, for example, opening an internal loop in the first term of Eq.~\eqref{eq:internal_loop1} results in the following four partially cut diagrams ${\tilde{c}}$,
\begin{equation}
\begin{aligned}
    \mathop{\adjincludegraphics[width=2.7cm, valign=c]{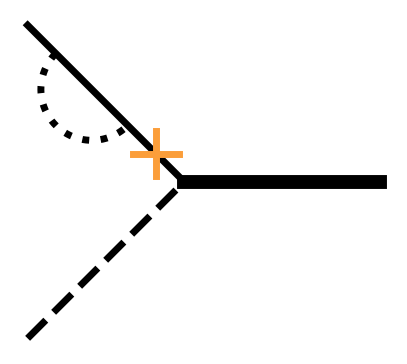}}
    &= 
    \mathop{\adjincludegraphics[width=2.7cm, valign=c]{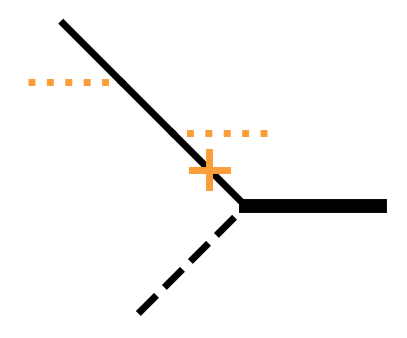}}
    +\mathop{\adjincludegraphics[width=2.7cm, valign=c]{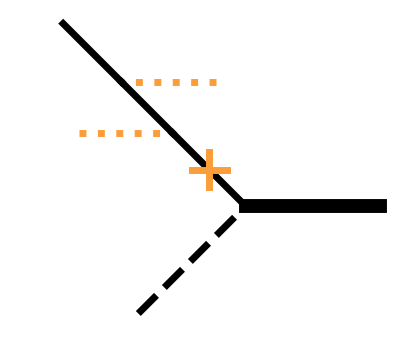}}
    +\mathop{\adjincludegraphics[width=2.7cm, valign=c]{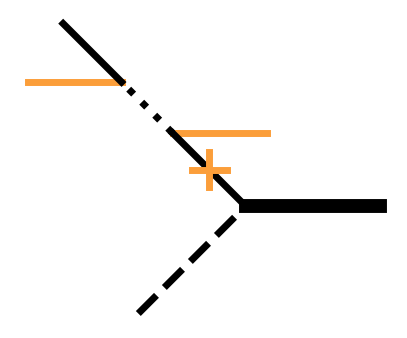}}
    +\mathop{\adjincludegraphics[width=2.7cm, valign=c]{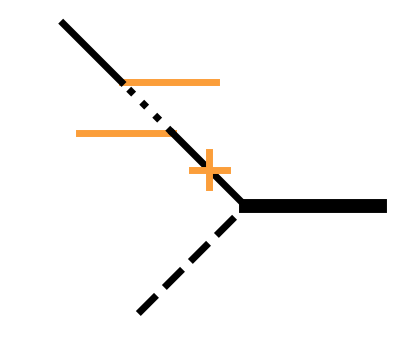}}.
    \label{eq:asym_cut_term}
\end{aligned}
\end{equation}
In general, a matrix element with internal loops can be written as 
\begin{align}
 \tilde{\M}^{c}= \sum_{\tilde{c}} \int d\mathrm{\Phi}_{\tilde{c}} \mathcal{F}_{\tilde{c}} 
 \tilde{\M}^{c,\tilde{c}}\,,
 \label{eq:cut_amp_betac}
\end{align}
where
\begin{align}
     d\mathrm{\Phi}_{\tilde{c}} = \prod_{j\in\mathcal{S}_{\tilde{c}}} \dkbar[4]{q_{j}}
     \frac{(-1)^{n_{j}-1}}{(n_{j} -1)!}\deltabar^{(+)(n_{j}-1)}\left(q^2_j - m^2_j\right)\,,
\end{align}
is the phase space factor for all spectator particles that have been put on-shell through the partial cut, denoted by $\mathcal{S}_{\tilde{c}}$,
and 
\begin{align}
    \mathcal{F}_{\tilde{c}} = \prod_{j\in \mathcal{S}_{\tilde{c}}}\left(\frac{1}{2}+\eta_j f(q^0_j)\right)
    \label{eq:Fspec}
\end{align}
\noindent is their thermal weight. We note that this thermal phase space distribution includes a temperature-independent part given by the factor of 1/2, which needs to be dropped when computing rates (see the discussion around Eq.~\eqref{eq:spectator_half}). This is similar to dropping the (diverging) temperature-independent contribution when calculating thermal masses under the assumption that it is accounted for by renormalizing the vacuum fields in the usual fashion~\cite{Das:1997gg}. 

As discussed above, the matrix element that accounts for any on-shell internal states (arising as a result of putting the \textit{spectator} on-shell) can be similarly written as
\begin{equation}
    \tilde{\M}^{c,\tilde{c}} = \M^{c,\tilde{c}} \times
    \prod_{j\in\mathcal{S}_{\tilde{c}}} 
    (q_j^2 - m_j^2)^{n_j-1}.
\end{equation}
Note that, in our specific example, putting the spectator on-shell does not result in an additional internal on-shell state and therefore the factor $\prod_{j\in\mathcal{S}_{\tilde{c}}} 
    (q_j^2 - m_j^2)^{n_j-1} $ is just unity. 
Since the partially cut diagrams
are tree-level, the amplitude $\mathcal{M}^{c,\,\tilde{c}}$ can be calculated by using the usual vacuum Feynman rules. For the diagrams in Eq.~\eqref{eq:asym_cut_term}, this gives us the amplitude, 
\begin{equation}
   \mathop{\adjincludegraphics[width=2.7cm, valign=c]{diagrams/Scalar-acut-0right.png}}
\begin{aligned}
    =&i g^2 \lambda^2
    \int \dkbar[4]{p_3} \deltabar^{(+)}(p_3^2-m_{\phi_3}^2) 
    \big\{ D({q} + {p_3})+D({q} - {p_3}) \big\} \qty({\textstyle \frac{1}{2}} + f_{p_3})\\
    &+i g^2 \lambda^2
    \int \dkbar[4]{p_1} \deltabar^{(+)}(p_1^2-m_{\phi_1}^2) 
    \big\{ D({q} + {p_1}) + D({q} - {p_1})\big\} \qty({\textstyle \frac{1}{2}} + f_{p_1})\\
\end{aligned}
\end{equation}
In general, the full expression for $[\tilde{\M}_{L}^{\alphac} \tilde{\M}_R^{\alphac}]$ appearing in Eq.~\eqref{eq:PiN_cut} can be written as,
\begin{equation}
\begin{aligned}
    \tilde{\M}_{L}^{\alphac} \tilde{\M}_R^{\alphac} = &\left(
    \sum_{\tilde{c}_L} \int d\mathrm{\Phi}_{\tilde{c}_L} \mathcal{F}_{\tilde{c}_L} 
    {\M}^{\alphac,\tilde{c}_L}
    \qty[\prod_{j_L\in\mathcal{S}_{\tilde{c}_L}} 
    (q_{j_L}^2 - m_{j_L}^2)^{n_{j_L}-1}]
    \right) \\
    &\times \left(
    \sum_{\tilde{c}_R} \int d\mathrm{\Phi}_{\tilde{c}_R} \mathcal{F}_{\tilde{c}_R} 
    {\M}^{\alphac,\tilde{c}_R}
    \qty[\prod_{j_R\in\mathcal{S}_{\tilde{c}_R}} 
    (q_{j_R}^2 - m_{j_R}^2)^{n_{j_R}-1}]
    \right)
    \times \prod_{i\in \mathcal{S}_\alphac}  
    (p_i^2 - m_i^2)^{n_i-1}
\end{aligned}
\label{eq:amp_full}
\end{equation}

\begin{step}\label{step4}\end{step}\textbf{\textit{Sum of tree-level processes}: }
By following the procedure described above, the $n$-loop self-energy diagram can be completely written in terms of tree-level processes. The total imaginary part of the self-energy at this loop order is then given by a sum over all these processes, 
\begin{equation}
\begin{aligned}
    \Im\Pi^{(n)}(k)
    = -\frac{1}{2} &\sum_{d,b}
    \sum_{\alphac,\tilde{c}_L,\tilde{c}_R}
    \int d\Phi_{\alphac} d\mathrm{\Phi}_{\tilde{c}_L} d\mathrm{\Phi}_{\tilde{c}_R} 
    \deltabar^{4}\left(k+\sum_{i}s^{\alphac}_{i}p_{i}\right)
    F_{\alphac} \mathcal{F}_{\tilde{c}_L} \mathcal{F}_{\tilde{c}_R} \\
    &\times\sum_{\mathrm{spins}}
    \qty[{\M}^{\alphac,\tilde{c}_L} {\M}^{\alphac,\tilde{c}_R}
    \qty(\prod_{i\in \mathcal{S}_\alphac}
    (p_i^2 - m_i^2)^{n_i-1}) \qty(\prod_{j\in\mathcal{S}_{\tilde{c}_L},\mathcal{S}_{\tilde{c}_R}} 
    (q_{j}^2 - m_{j}^2)^{n_{j}-1})].
\end{aligned}
\label{eq:PiN_total}
\end{equation}
It is therefore possible to completely define the scattering rate for any particle in a medium using Eq.~\eqref{eq:PiN_total}. For the 2-loop example under consideration, this results in,
\begin{equation}\label{ImPi_allscalar_rules}
\begin{aligned}
    \Im\Pi(\omega, \vb{k}) =\\ 
    -\frac{ g^2 \lambda^2 }{2}
    &\int \dkbar[4]{p_2} \deltabar^{(+)}(({p_2^0})^2-E_{p_2}^2)
    \int \dkbar[4]{p_3} \deltabar^{(+)}(({p_3^0})^2-E_{p_3}^2)
    \int \dkbar[4]{p_1} \deltabar^{(+)}(({p_1^0})^2-E_{p_1}^2)
    \\
    \Big[ \ 
    &\deltabar^{(4)}(k + {p_2} + {p_3} - {p_1})\qty[D(k + {p_2})]^2 \qty[f_{p_2} f_{p_3} (1+f_{p_1}) - f_{p_1}(1+f_{p_2})(1+f_{p_3})] \quad\\
    +&\deltabar^{(4)}(k + {p_2} - {p_3} + {p_1}) \qty[D(k + {p_2})]^2 \qty[f_{p_2} f_{p_1} (1+f_{p_3}) - f_{p_3} (1+f_{p_1}) (1+f_{p_2})]\\
    +&\deltabar^{(4)}(k - {p_2} + {p_3} + {p_1}) \qty[D(k - {p_2})]^2 \qty[f_{p_3} f_{p_1} (1+f_{p_2}) - f_{p_2} (1+f_{p_3}) (1+f_{p_1})]\\
    +&\deltabar^{(4)}(k - {p_2} - {p_3} + {p_1}) \qty[D(k - {p_2})]^2 \qty[f_{p_1} (1+f_{p_2}) (1+f_{p_3}) - f_{p_2} f_{p_3} (1+f_{p_1})]\\
    +&\deltabar^{(4)}(k - {p_2} + {p_3} - {p_1}) \qty[D(k - {p_2})]^2 \qty[f_{p_3} (1+f_{p_2}) (1+f_{p_1}) - f_{p_2} f_{p_1} (1+f_{p_3})]\\
    +&\deltabar^{(4)}(k + {p_2} - {p_3} - {p_1}) \qty[D(k + {p_2})]^2 \qty[f_{p_2} (1+f_{p_3}) (1+f_{p_1}) - f_{p_3} f_{p_1} (1+f_{p_2})]\\
    +&\deltabar^{(4)}(k - {p_2} - {p_3} - {p_1}) \qty[D(k - {p_2})]^2 \qty[(1+f_{p_2}) (1+f_{p_3}) (1+f_{p_1}) - f_{p_2} f_{p_3} f_{p_1}]\\
    +&\deltabar^{(4)}(k + {p_2} + {p_3} + {p_1}) \qty[D(k + {p_2})]^2 \qty[f_{p_2} f_{p_3} f_{p_1} - (1+f_{p_2}) (1+f_{p_3}) (1+f_{p_1})]
    \ \Big]\\
    + \frac{g^2 \lambda^2}{2}
    &\int \dkbar[4]{p_2} \deltabar^{(+)}(({p_2^0})^2-E_{p_2}^2)
     \int \dkbar[4]{p_3} \deltabar^{(+)}(({p_3^0})^2-E_{p_3}^2)
     \int \dkbar[4]{q} \deltabar^{(+)'}(({q^0})^2-E_{q}^2) 
    \\
    \Big[ \
     &\deltabar^{(4)}(k + {p_2} - {q}) \big\{ D({q} + {p_3})+D({q} - {p_3}) \big\} \qty({\textstyle \frac{1}{2}} + f_{p_3}) [f_{p_2} (1+f_{q}) - f_{q}( 1+f_{p_2})]\\
    +&\deltabar^{(4)}(k - {p_2} + {q}) \big\{ D({q} + {p_3})+D({q} - {p_3}) \big\} \qty({\textstyle \frac{1}{2}} + f_{p_3}) [f_{q} (1+f_{p_2}) - f_{p_2} (1+f_{q})]\\
    +&\deltabar^{(4)}(k - {p_2} - {q}) \big\{ D({q} + {p_3})+D({q} - {p_3}) \big\} \qty({\textstyle \frac{1}{2}} + f_{p_3}) [(1+f_{p_2}) (1+f_{q}) - f_{p_2} f_{q}]\\
    +&\deltabar^{(4)}(k + {p_2} + {q}) \big\{ D({q} + {p_3})+D({q} - {p_3}) \big\} \qty({\textstyle \frac{1}{2}} + f_{p_3}) [f_{p_2} f_{q} - (1+f_{p_2}) (1+f_{q})]
    \ \Big]\\
    +  \frac{g^2 \lambda^2}{2}
    &\int \dkbar[4]{p_2} \deltabar^{(+)}(({p_2^0})^2-E_{p_2}^2)
     \int \dkbar[4]{p_1} \deltabar^{(+)}(({p_1^0})^2-E_{p_1}^2) 
     \int \dkbar[4]{q} \deltabar^{(+)'}(({q^0})^2-E_{q}^2)
    \\
    \Big[ \
     &\deltabar^{(4)}(k + {p_2} - {q}) \big\{ D({q} + {p_1}) + D({q} - {p_1})\big\} \qty({\textstyle \frac{1}{2}} + f_{p_1}) [f_{p_2} (1+f_{q}) - f_{q} (1+f_{p_2})]\\
    +&\deltabar^{(4)}(k - {p_2} + {q}) \big\{ D({q} + {p_1}) + D({q} - {p_1})\big\} \qty({\textstyle \frac{1}{2}} + f_{p_1}) [f_{q} (1+f_{p_2}) - f_{p_2} (1+f_{q})]\\
    +&\deltabar^{(4)}(k - {p_2} - {q}) \big\{ D({q} + {p_1}) + D({q} - {p_1})\big\} \qty({\textstyle \frac{1}{2}} + f_{p_1}) [(1+f_{p_2}) (1+f_{q}) - f_{p_2} f_{q}]\\
    +&\deltabar^{(4)}(k + {p_2} + {q}) \big\{ D({q} + {p_1}) + D({q} - {p_1})\big\} \qty({\textstyle \frac{1}{2}} + f_{p_1}) [f_{p_2} f_{q} - (1+f_{p_2}) (1+f_{q})]
    \ \Big]\\
\end{aligned}
\end{equation}

\subsection{Caveats for SSHB}
There are a few subtleties one has to keep in mind while following the procedure outlined above:
\begin{itemize}
    \item The ITF only applies to systems that are in thermal equilibrium. In particular, the formalism works if the particles in the loop have a Fermi-dirac or Bose-Einstein distribution. As such, the self-energy diagram can have the test particle $\Phi$ running inside the loop as long as the distribution of $\Phi$ is Bose-Einstein or Fermi Dirac. For the example considered above, this also means that we assume that $\phi_2$
    is kept in equilibrium because of some other process not captured by the Lagrangian of Eq.~\eqref{eq:scalar_lag}.
    \item An $n$-loop self energy can have multiple topologies, depending on the Lagrangian under consideration. For example, if we allow for an interaction vertex of the type $\lambda_{2}\phi_1\phi_2\phi_3$ in the example considered above, we can additionally draw the following topology for a two-loop diagram,
    \begin{equation}
    \adjincludegraphics[height=1.5cm,valign=c]{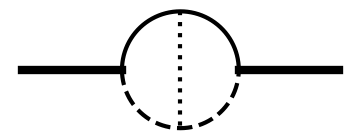}.
    \end{equation}
    A bisection through such a topology necessarily results in amplitudes for which $\mathcal{M}_L\neq\mathcal{M}_R^*$, and which at first sight appears to correspond to an interference contribution. However, a qualitative distinction has to be made between a bisection that results in no internal loops versus one which does. The former simply provides the interference term between the usual interfering vacuum amplitudes, such as the $s$-channel and $t$-channel diagrams, 
    \begin{equation}
    \adjincludegraphics[height=1.95cm,valign=c]{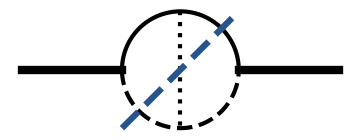} = \adjincludegraphics[height=1.95cm,valign=c]{diagrams/Scalar-cut2to2s-0213.png} \otimes \adjincludegraphics[height=1.95cm,valign=c]{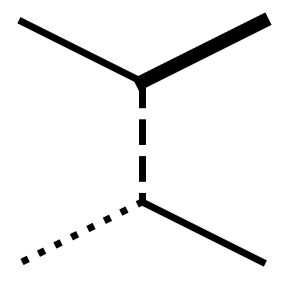}
    \end{equation}
    whereas the latter results in the novel interference diagrams discussed in this work which are really the interference between $n\to m$ and a $n+\ell \to m+\ell$ processes and which only arise at FTD.
    \item If the external state particle $\Phi$ is a vector or a fermion, its self energy (denoted by $\Pi^{\mu\nu}$ with spacetime indices $\mu,\,\nu$, or $\Sigma^{s,s^\prime}$ with spinor indices $s,\,s^\prime$) 
    needs to be multiplied with the associated polarization which is then summed over in order to get a production rate. This implies
    \begin{align}
        -\omega \Gamma_\Phi &= \sum_{\mu\nu}\epsilon_\mu \,\mathrm{Im}\Pi^{\mu\nu}\epsilon_\nu,\,\qquad \Phi=\mathrm{Vector} \\
        -\omega \Gamma_\Phi &= \sum_{s,s'}\bar{u}_s \,\mathrm{Im}\Sigma^{s,s'} \,u_{s'},\qquad \Phi=\mathrm{Fermion}
    \end{align}
    where $\mathrm{Im}\Pi^{\mu\nu}$ and $\mathrm{Im}\Sigma^{s,s'}$ 
    are calculated using Eq.~\eqref{eq:PiN_total}. 
    \item Since summing over the different diagrams arising out of bisections and partial cuts necessarily involves summing over \textit{amplitudes}, the sign of individual amplitudes is crucial. In particular, with fermion states running in the loop, one has to be careful to account for the signs when considering interactions with particles and antiparticles.
    \item We always work in the regime where the thermal mass of the particles running in the loop can be ignored. In effect, this means that $m_{\phi_i} \gtrsim m_{\phi_i}(T) \sim g T$ where $g$ is the coupling responsible for keeping the particles in equilibrium. At larger temperatures, or smaller masses, the ITF propagators need to be resummed~\cite{Pisarski:1988vd,Pisarski1990,Braaten:1989mz,Frenkel:1989br}. We leave the incorporation of these effects into SSHB as a direction for future work. We note however that our prescription will be valid for any massless particle provided that its thermally induced mass is purely a function of temperature and/or chemical potential (and not of its four-momentum). In this case, the poles of the in-medium propagators will shift by a constant amount and the prescription is valid with the simple replacement $m_{\phi_i} \to m_{\phi_i}(T,\,\mu)$. As an example, our formalism can incorporate photons in the loop in nonrelativistic and nondegenerate plasmas where the on-shell photon thermal mass can be written purely in terms of $T$ and $\mu$ \cite{Braaten:1993jw}.
\end{itemize}

\section{Corroborating the rules for SSHB with finite temperature field theory}\label{sec:corroborate}

\begin{figure}[ht]
    \centering
    \includegraphics[width=0.8\linewidth]{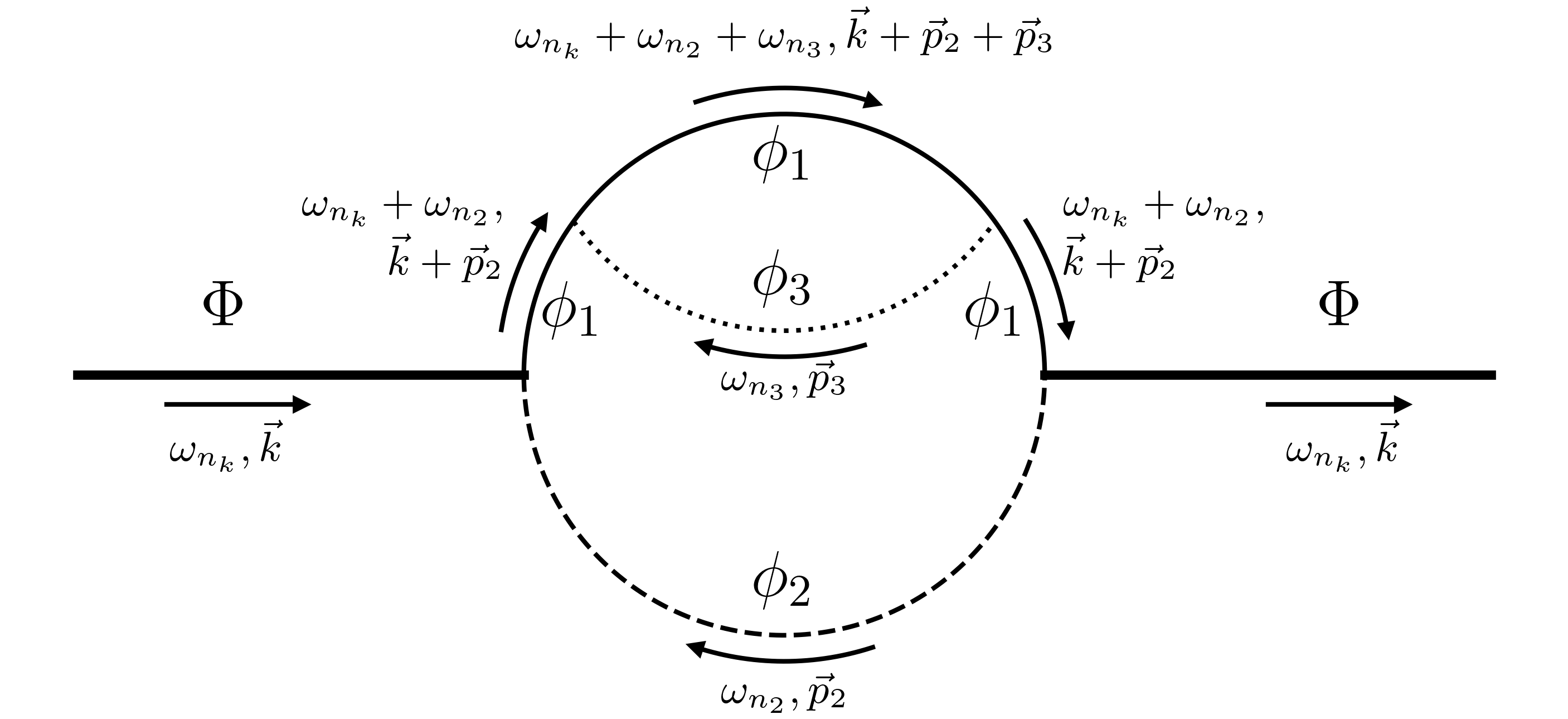}
    \caption{Two-loop self-energy in an all scalar toy model. 
    } 
    \label{fig:2loopFeynman}
\end{figure}
\label{sec:CutComplicated}
In the previous Section, we computed the thermal interaction rate in a toy model using the SSHB prescription. In this Section, we present the full thermal field theory calculation for the two-loop self-energy diagram in Eq.~\eqref{eq:2-loop-scalar} (reproduced in \Fig{fig:2loopFeynman}), to demonstrate how the SSHB rules arise more rigorously. Note that the one-loop contribution to the self-energy is given by Eq.~\eqref{eq:ImPiScalar1loop_diag} and it is straightforward to see how the SSHB rules generate this expression.

For the two-loop self-energy diagram with the momenta assigned as in \Fig{fig:2loopFeynman}, the ITF yields the following expression for the self-energy,
\begin{equation}
\begin{aligned}
    -i \Pi(\omega_{n_k}, \vb{k})= & \,iT \sum_{n_2=-\infty}^{\infty} iT  \sum_{n_3=-\infty}^{\infty} \int \dkbar[3]{p_2} \dkbar[3]{p_3} (-ig)^2 (-i\lambda)^2 \qty[iD(\omega_{n_k} + \omega_{n_2},\vb{k} + \vb{p}_2)]^2 \\
    &\qquad \quad \times iD(\omega_{n_k} + \omega_{n_2} + \omega_{n_3},\vb{k} + \vb{p}_2 + \vb{p}_3) iD(\omega_{n_3},\vb{p}_3) iD(\omega_{n_2},\vb{p}_2)\,
    \label{eq:ImPiomega_n}
\end{aligned}
\end{equation}
where $\omega_n = i 2\pi n T$ are Bosonic Matsubara frequencies for $n = (n_k,\, n_2,\, n_3)$ and
\begin{align}
iD(\omega,\vb{p}) &= \frac{i}{\omega^2-\vb{p}^2-m_{\phi_i}^2}\,,
\end{align}
are the propagators for the particles $\phi_i$ with momentum $\vb{p}$ and complex energy $\omega$ defined through Fig.~\ref{fig:2loopFeynman}. Eq.~\eqref{eq:ImPiomega_n} can be simplified to
\begin{equation}
    \begin{aligned}
    \Pi(\omega_{n_k},\vb{k}) = g^2 \lambda^2 &\int \dkbar[3]{p_2}  T \sum_{n_2} 
    \qty[D(\omega_{n_k} + \omega_{n_2},\vb{k} + \vb{p}_2)]^2 D(\omega_{n_2},\vb{p}_2) \\
    \times &\int \dkbar[3]{p_3} T \sum_{n_3} D(\omega_{n_k} + \omega_{n_2} + \omega_{n_3},\vb{k} + \vb{p}_2 + \vb{p}_3) D(\omega_{n_3},\vb{p}_3).
    \label{eq:Pi2loop_full}
    \end{aligned}
\end{equation}
The infinite sums over $n_2$ and $n_3$ can be converted into a sum over the residues of the integrand using the identity \cite{Kapusta:2006pm},
\begin{equation}
       T\sum_{n=-\infty}^{\infty} M(p_0 = \omega_n^B+\mu) = - \sum_{p_0 : \,M^{-1}(p_0) = 0} \mathrm{Res} \qty{M(p_0) \frac{1}{2} \coth \left(\frac{p_0-\mu}{2T}\right)}\,,
       \label{eq:res_iden}
\end{equation}
where the different $p_0$'s correspond to the poles of the propagators appearing in Eq.~\eqref{eq:Pi2loop_full}. For multi-loop diagrams, these can be either simple poles (if the propagator is not repeated) or double or higher-order poles (if it is repeated). For a pole of the form, $p_0 = k_0 \pm E_{p-k}$ at order $n+1$, arising from having an integrand of the form $D(p_0-k_0,\vb{p}-\vb{k})]^{n+1} \ g(p)$ where $g(p)$ is some non-singular function of the four-momentum, the sum over the residues can be converted into an integral using the identity (see App.~\ref{app:res_iden}),
\begin{equation}
\begin{aligned}
    \int \dkbar[3]{p} \sum_{p_0 = k_0 \pm E_{p-k}} &\text{Res} \qty{ \qty[D(p_0-k_0,\vb{p}-\vb{k})]^{n+1} \ g(p)} \\
    &= \int \dkbar[4]{q} \frac{(-1)^{n}}{n!} \deltabar^{(+)(n)}(q_0^2-E_q^2) \Big[g(p = k + q) - g(p = k - q)\Big]\,,
    \label{eq:res_simplify}
\end{aligned}
\end{equation}
where $E_{p-k}^2 = (\vb{p}-\vb{k})^2 + m^2$, $E_{q}^2 = \vb{q}^2 + m^2$, and $\deltabar^{(+)(n)}=(2\pi)\delta^{(+)(n)}$ is given by,
\begin{equation}
    \frac{(-1)^{n}}{n!} \delta^{(+)(n)}(q_0^2-E_q^2) = \frac{(-1)^{n}}{n!} \theta(q_0) \dv[n]{(q_0^2)} \delta(q_0^2-E_q^2) = \theta(q_0) \frac{1}{(q_0^2-E_q^2)^n}\delta(q_0^2-E_q^2)\,. 
\end{equation}
For simple poles ($n=0$), this results in the usual expression with a $\delta$--function that puts the corresponding propagators on-shell. For higher-order poles, we instead obtain derivatives of the $\delta$--function, denoted by  $\delta^{(+){(n)}}$ where the $(+)$ signifies the fact that we are choosing the positive energy pole.
These higher-order poles arise when we are simultaneously putting multiple particles on-shell (denoted by the orange x in the diagrams from the previous Section).

Using Eqs.~\eqref{eq:res_iden} and~\eqref{eq:res_simplify}, and after a fair bit of algebra, we can simplify Eq.~\eqref{eq:Pi2loop_full} to obtain the full expression for the $\Phi$ self-energy as a function of its four-momentum $k\equiv(\omega,\,\vb{k})$, 
\begin{equation}
\begin{aligned}
    \Pi(\omega, \vb{k}) = 
    g^2 \lambda^2 &
    \int \dkbar[4]{{p_3}} \deltabar^{(+)}(({p_3^0})^2-E_{p_3}^2) 
    \int \dkbar[4]{{p_1}} \deltabar^{(+)}(({p_1^0})^2-E_{p_1}^2) \\
    &\left[ \left[
    D(k+{p_3}-{p_1}) \qty[D(-{p_3} + {p_1})]^2 
    \ch{p_3} \ch{{p_1}-{p_3}}
    - ({p_1} \to - {p_1})\right] - ({p_3} \to - {p_3})\right] \\
    + g^2 \lambda^2 &
    \int \dkbar[4]{{p_3}} \deltabar^{(+)}(({p_3^0})^2-E_{p_3}^2)
    \int \dkbar[4]{{p_2}} \deltabar^{(+)}(({p_2^0})^2-E_{p_2}^2)\\
    &\left[ \left[
    D(k+ {p_2} + {p_3}) \qty[D(k + {p_2})]^2 
    \ch{p_3} \ch{p_2}
    - ({p_2} \to - {p_2}) \right]- ({p_3} \to - {p_3})\right]\\
    - g^2 \lambda^2 &
    \int \dkbar[4]{{p_3}} \deltabar^{(+)}(({p_3^0})^2-E_{p_3}^2)
    \int \dkbar[4]{{q}} \deltabar ^{(+)'}(({q^0})^2-E_{q}^2) \\
    &\left[\left[
    D(k-{q}) D({q} +{p_3}) 
    \ch{p_3} \ch{q}
    - ({q} \to - {q})\right]- ({p_3} \to - {p_3})\right]\\
    +g^2 \lambda^2&
    \int \dkbar[4]{{p_1}} \deltabar^{(+)}(({p_1^0})^2-E_{p_1}^2)
    \int \dkbar[4]{{p_3}} \deltabar^{(+)}(({p_3^0})^2-E_{p_3}^2) \\
    &\left[\left[
    D(k - {p_3} - {p_1}) \qty[D({p_3} + {p_1})]^2 
    \ch{p_1} \ch{{p_3} + {p_1}}
    - ({p_3} \to - {p_3})\right]- ({p_1} \to - {p_1})\right]\\
    + g^2 \lambda^2&
    \int \dkbar[4]{{p_1}} \deltabar^{(+)}(({p_1^0})^2-E_{p_1}^2)
    \int \dkbar[4]{{p_2}}\deltabar^{(+)}(({p_2^0})^2-E_{p_2}^2) \\
    &\left[\left[
    D(k+{p_2}-{p_1}) \qty[D(k + {p_2})]^2 
    \ch{p_1} \ch{p_2}
    - ({p_2} \to - {p_2})\right]- ({p_1} \to - {p_1})\right]\\
    - g^2 \lambda^2&
    \int \dkbar[4]{{p_1}} \deltabar^{(+)}(({p_1^0})^2-E_{p_1}^2)
    \int \dkbar[4]{{q}} \deltabar^{(+)'}(({q^0})^2-E_{q}^2)\\
    &\left[\left[
    D(k -{q}) D({q}-{p_1}) 
    \ch{p_1} \ch{q}
    - ({q} \to - {q})\right]- ({p_1} \to - {p_1})\right]\\
\end{aligned}
\end{equation}
where $\mathbf{q}$
(or $\mathbf{p}_1$) is the momentum carried by $\phi_1$ if the $\phi_1$ propagator is repeated (or not repeated), $\ch{p} = \chfull{p^0}$ and $D(p)$ is shorthand for $D(p^0, \vb{p})$. 
Note that the $\coth$ functions, $c_p$, are related to the equilibrium Bose-Einstein distribution functions, $f_B$, through the identity, $\coth(z) = 1 + 2 f_\text{B}(2z)$. If instead, we were summing over fermionic Matsubara frequencies in Eq.~\eqref{eq:Pi2loop_full}, we would have obtained a set of $\tanh$ function which are related to the Fermi-Dirac equilibrium distribution functions, $f_F$, through $\tanh(z) = 1 - 2 f_\text{F}(2z)$. It is from this correspondence between the trigonometric and the quantum distributions functions that the dependence of the self energy on the ambient background density is made explicit.

The imaginary part of the self energy is related to its discontinuity, $\mathrm{Im}\Pi = \mathrm{Disc}\Pi/(2i)$, where $\Disc{\Pi(\omega, \vb{k})} = \lim_{\epsilon \to 0} \qty[\Pi(\omega + i\epsilon, \vb{k}) - \Pi(\omega - i\epsilon, \vb{k})]$. For each propagator in the integrand that involves the external energy $\omega$, the discontinuity for $\Pi \sim \qty[D(\omega-p_0,\vb{k}-\vb{p})]^{n+1}$ is given by (see App. \ref{app:discon_iden}),
\begin{equation}
    \mathrm{Disc} \ \Pi \nonumber \sim -i \int \dkbar[4]{q} \frac{(-1)^{n}}{n!} \deltabar^{(+)(n)}(q_0^2-E_q^2) \Big[\deltabar^{4}(k - p - q) - \deltabar^{4}(k - p + q)\Big]\,.
    \label{eq:disc_simplify_identity}
\end{equation}
Using this identity and replacing the trigonometric functions with the particle phase space distributions, we can calculate the imaginary part of the self-energy,
\begin{equation}
\begin{aligned}
    \Im\Pi(\omega, \vb{k}) =\\ 
    -\frac{ g^2 \lambda^2 }{2}
    &\int \dkbar[4]{p_2} \deltabar^{(+)}(({p_2^0})^2-E_{p_2}^2)
    \int \dkbar[4]{p_3} \deltabar^{(+)}(({p_3^0})^2-E_{p_3}^2)
    \int \dkbar[4]{p_1} \deltabar^{(+)}(({p_1^0})^2-E_{p_1}^2)
    \\
    \Big[ \ 
    &\deltabar^{(4)}(k + {p_2} + {p_3} - {p_1})\qty[D(k + {p_2})]^2 \qty[f_{p_2} f_{p_3} (1+f_{p_1}) - f_{p_1}(1+f_{p_2})(1+f_{p_3})] \quad\\
    +&\deltabar^{(4)}(k + {p_2} - {p_3} + {p_1}) \qty[D(k + {p_2})]^2 \qty[f_{p_2} f_{p_1} (1+f_{p_3}) - f_{p_3} (1+f_{p_1}) (1+f_{p_2})]\\
    +&\deltabar^{(4)}(k - {p_2} + {p_3} + {p_1}) \qty[D(k - {p_2})]^2 \qty[f_{p_3} f_{p_1} (1+f_{p_2}) - f_{p_2} (1+f_{p_3}) (1+f_{p_1})]\\
    +&\deltabar^{(4)}(k - {p_2} - {p_3} + {p_1}) \qty[D(k - {p_2})]^2 \qty[f_{p_1} (1+f_{p_2}) (1+f_{p_3}) - f_{p_2} f_{p_3} (1+f_{p_1})]\\
    +&\deltabar^{(4)}(k - {p_2} + {p_3} - {p_1}) \qty[D(k - {p_2})]^2 \qty[f_{p_3} (1+f_{p_2}) (1+f_{p_1}) - f_{p_2} f_{p_1} (1+f_{p_3})]\\
    +&\deltabar^{(4)}(k + {p_2} - {p_3} - {p_1}) \qty[D(k + {p_2})]^2 \qty[f_{p_2} (1+f_{p_3}) (1+f_{p_1}) - f_{p_3} f_{p_1} (1+f_{p_2})]\\
    +&\deltabar^{(4)}(k - {p_2} - {p_3} - {p_1}) \qty[D(k - {p_2})]^2 \qty[(1+f_{p_2}) (1+f_{p_3}) (1+f_{p_1}) - f_{p_2} f_{p_3} f_{p_1}]\\
    +&\deltabar^{(4)}(k + {p_2} + {p_3} + {p_1}) \qty[D(k + {p_2})]^2 \qty[f_{p_2} f_{p_3} f_{p_1} - (1+f_{p_2}) (1+f_{p_3}) (1+f_{p_1})]
    \ \Big]\\
    + \frac{g^2 \lambda^2}{2}
    &\int \dkbar[4]{p_2} \deltabar^{(+)}(({p_2^0})^2-E_{p_2}^2)
     \int \dkbar[4]{p_3} \deltabar^{(+)}(({p_3^0})^2-E_{p_3}^2)
     \int \dkbar[4]{q} \deltabar^{(+)'}(({q^0})^2-E_{q}^2) 
    \\
    \Big[ \
     &\deltabar^{(4)}(k + {p_2} - {q}) \big\{ D({q} + {p_3})+D({q} - {p_3}) \big\} \qty({\textstyle \frac{1}{2}} + f_{p_3}) [f_{p_2} (1+f_{q}) - f_{q}( 1+f_{p_2})]\\
    +&\deltabar^{(4)}(k - {p_2} + {q}) \big\{ D({q} + {p_3})+D({q} - {p_3}) \big\} \qty({\textstyle \frac{1}{2}} + f_{p_3}) [f_{q} (1+f_{p_2}) - f_{p_2} (1+f_{q})]\\
    +&\deltabar^{(4)}(k - {p_2} - {q}) \big\{ D({q} + {p_3})+D({q} - {p_3}) \big\} \qty({\textstyle \frac{1}{2}} + f_{p_3}) [(1+f_{p_2}) (1+f_{q}) - f_{p_2} f_{q}]\\
    +&\deltabar^{(4)}(k + {p_2} + {q}) \big\{ D({q} + {p_3})+D({q} - {p_3}) \big\} \qty({\textstyle \frac{1}{2}} + f_{p_3}) [f_{p_2} f_{q} - (1+f_{p_2}) (1+f_{q})]
    \ \Big]\\
    +\frac{g^2 \lambda^2}{2}
    &\int \dkbar[4]{p_2} \deltabar^{(+)}(({p_2^0})^2-E_{p_2}^2)
     \int \dkbar[4]{p_1} \deltabar^{(+)}(({p_1^0})^2-E_{p_1}^2) 
     \int \dkbar[4]{q} \deltabar^{(+)'}(({q^0})^2-E_{q}^2)
    \\
    \Big[ \
     &\deltabar^{(4)}(k + {p_2} - {q}) \big\{ D({q} + {p_1}) + D({q} - {p_1})\big\} \qty({\textstyle \frac{1}{2}} + f_{p_1}) [f_{p_2} (1+f_{q}) - f_{q} (1+f_{p_2})]\\
    +&\deltabar^{(4)}(k - {p_2} + {q}) \big\{ D({q} + {p_1}) + D({q} - {p_1})\big\} \qty({\textstyle \frac{1}{2}} + f_{p_1}) [f_{q} (1+f_{p_2}) - f_{p_2} (1+f_{q})]\\
    +&\deltabar^{(4)}(k - {p_2} - {q}) \big\{ D({q} + {p_1}) + D({q} - {p_1})\big\} \qty({\textstyle \frac{1}{2}} + f_{p_1}) [(1+f_{p_2}) (1+f_{q}) - f_{p_2} f_{q}]\\
    +&\deltabar^{(4)}(k + {p_2} + {q}) \big\{ D({q} + {p_1}) + D({q} - {p_1})\big\} \qty({\textstyle \frac{1}{2}} + f_{p_1}) [f_{p_2} f_{q} - (1+f_{p_2}) (1+f_{q})]
    \ \Big]\\
    \label{eq:ImPi_allscalar_full}
\end{aligned}
\end{equation}
This is exactly the formula one obtains in \Eq{ImPi_allscalar_rules} by using the SSHB prescription detailed in \Sec{sec:SSHBrules}. Through this exercise of calculating the full expression using the ITF, it is now straightforward to see how the various factors in Eq.~\eqref{ImPi_allscalar_rules} arise at arbitrary loop order. In particular, we note that for symmetric bisections where all the cut propagators have simple poles, we always obtain a contribution independent of $\delta^{(+)'}$ with the net rate sensitive to the background particle density in the usual way. For example, the $\delta^{(+)}$--functions in the first term in the expression above put $\phi_{1,2,3}$ on shell which result in all possible processes that can produce or absorb a $\Phi$ through interactions involving all three particles, $\phi_{1},\,\phi_{2},\,\mathrm{and}\,\phi_3$. For asymmetric bisections, however, the presence of a $\delta^{(+)'}$ points toward the presence of internal thermal loops. The resulting processes are interference-type processes which correct the amplitude of an $n\to m$ process through forward scattering with background fields $\ell$. This can be seen from the second and third terms of the expression above. For example, the two factors of $\delta^{(+)}$ in the second term put $\phi_{2}$ and $\phi_3$ on-shell and result in all processes that produce and absorb a $\Phi$ through interactions with $\phi_2$ and $\phi_3$. These will be proportional to the thermal phase space of decays and inverse decays, which is exactly given by the phase space factors in the square brackets of this term. These decay and inverse decay amplitudes are however corrected by forward scattering with $\phi_1$ captured by the factor of $\delta^{(+)'}$. 
For a particle $\ell$ that forward scatters, the phase space distribution can be understood as 
\begin{align}
\left[ \substack{\text{emitting and} \\ \text{absorbing the particle} \\ \text{with an energy } q} \right]
-
\left[ \substack{\text{absorbing and} \\ \text{emitting the particle} \\ \text{with energy } q }\right]
\label{eq:spectator_half}
\end{align}
The former is given by the operator $aa^\dagger |f_\ell(q)\rangle = (1+ f_\ell(q))|f_\ell(q)\rangle$ and the latter by $a^\dagger a|f_\ell(q)\rangle = f_\ell(q)|f_\ell(q)\rangle$ resulting in a net phase space which can be simplified to $(1/2 + f_\ell)$ in the expressions above (see also Ref.~\cite{Majumder:2001iy}).  

\section{Application}\label{sec:application}
A key goal of this work is to quantify the effect of interference diagrams that may arise in a medium. We demonstrate this in the context of particle production in a generic plasma and note that this approach can be easily extended to the early universe and to astrophysical plasmas.

As shown in Sec.~\ref{sec:prod_rate}, the production rate per volume of $\Phi$ can be calculated by solving the Bolztmann equation, Eq.~\eqref{eq:rateImPi}.
If $\Phi$ has sub-thermal phase space density, 
i.e., $f_\Phi \ll f^\mathrm{eq}_\Phi$,
\begin{equation}
    \frac{\mathrm{d}n_\Phi}{\mathrm{d}t} = - \int \frac{\dkbar[3]{\vb{k}}}{2\omega} 2 f_\Phi^\mathrm{eq}(\omega) \ \mathrm{Im}\,\Pi_\Phi(\omega,\vb{k})
\end{equation}
The right hand side of this equation can be further simplified by considering the different processes that contribute to $\mathrm{Im}\,\Pi_\Phi$, as well as by assuming Maxwell-Boltzmann distributions, $f(\omega) = e^{-\omega/T}$ for $\Phi$ and $f(E_i) = e^{-(E_i - \mu_i)/T}$ for particle $i$. 

For scattering-type processes of the kind $\Phi,X_2 \to X_3,X_4$, where $X_i$ denote generic particles that $\Phi$ interacts with, and with the standard assumptions of calculating thermally averaged cross-sections \cite{Gondolo:1990dk}, we obtain,
\begin{align}
    \frac{\mathrm{d}n_\Phi}{\mathrm{d}t}\eval_{2 \to 2} &= \frac{1}{16(2\pi)^5} \ e^{\mu_2/T} \int_{s_0}^\infty \dd{s} \frac{T}{\sqrt{s}} K_1(\sqrt{s}/T) \int_{t_1}^{t_0} \dd{t} \abs{\mathcal{M}}^2
\end{align}
where $s_0 = \max[(m_\Phi+m_2)^2,(m_3+m_4)^2]$ 
is the minimal value for the center-of-momentum energy-squared, $K_1(z)$ is the first-order modified Bessel function of the second kind, $\mu_i$ is the chemical potential of $i$, and
\begin{equation}
    t_1(t_0) = \qty(E_{1,\mathrm{CM}}-E_{3,\mathrm{CM}})^2 - (\abs{\vb{p}_{1,\mathrm{CM}}} \pm \abs{\vb{p}_{3,\mathrm{CM}}})^2 \, .
\end{equation}

Similarly, for
decay-type processes of the kind $\Phi \to X_3,X_4$,
we obtain,
\begin{align}
    \frac{\mathrm{d}n_\Phi}{\mathrm{d}t}\eval_{1 \to 2} &= \frac{1}{4(2\pi)^4} \int_{m_\Phi}^\infty \dd{\omega} e^{-\omega/T} \int_{E_{3,\mathrm{min}}}^{E_{3,\mathrm{max}}} \dd{E_3} \int_{0}^{2\pi} \dd{\phi} \abs{\mathcal{M}}^2 \nonumber\\
    &= \frac{1}{4(2\pi)^3} m_\Phi T \beta_{34}(m_\Phi^2) K_1(m_\Phi/T) \abs{\mathcal{M}}^2
    \label{eq:1to2simple}
\end{align}
where,
\begin{equation}
    E_{3_{\mathrm{min}}^{\mathrm{max}}} = \frac{\omega}{2}\qty(1+\frac{m_3^2-m_4^2}{m_\Phi^2}) \pm \frac{k}{2} \beta_{34}(m_\Phi^2)\,,
\end{equation}
and,
\begin{equation}
    \beta_{ij}(s) = \sqrt{1 - \frac{2(m_i^2 + m_j^2)}{s} + \frac{(m_i^2 - m_j^2)^2}{s^2}}\,.
\end{equation}
Note that the second line of Eq.~\eqref{eq:1to2simple} holds only when $\abs{\mathcal{M}}^2$ is independent of the energies of the interacting particles.

\begin{figure}
    \centering
    \begin{subfigure}
    \centering \includegraphics[width=0.25\textwidth]{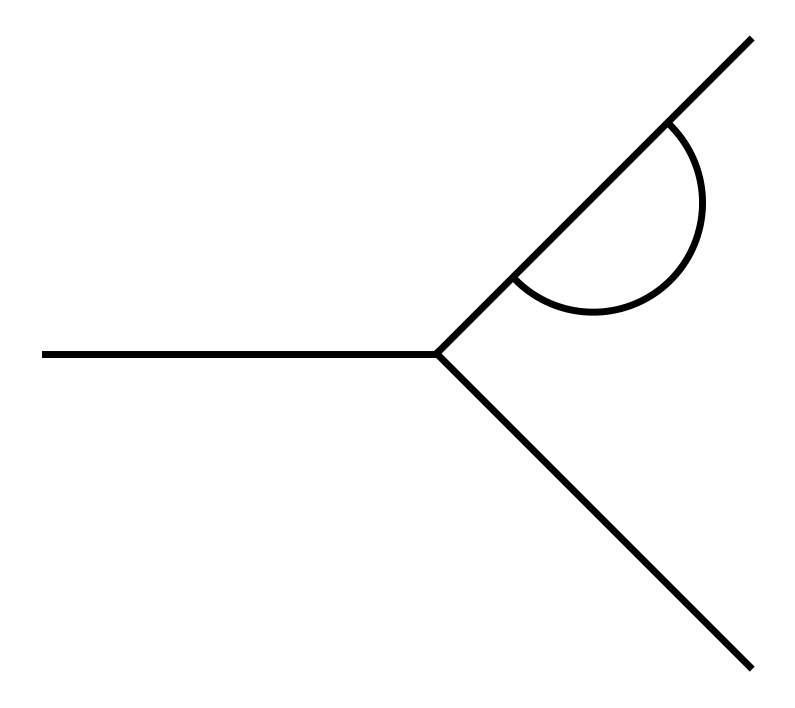}
    \end{subfigure}
    \hspace{0.1\textwidth}
    \begin{subfigure}
    \centering \includegraphics[width=0.25\textwidth]{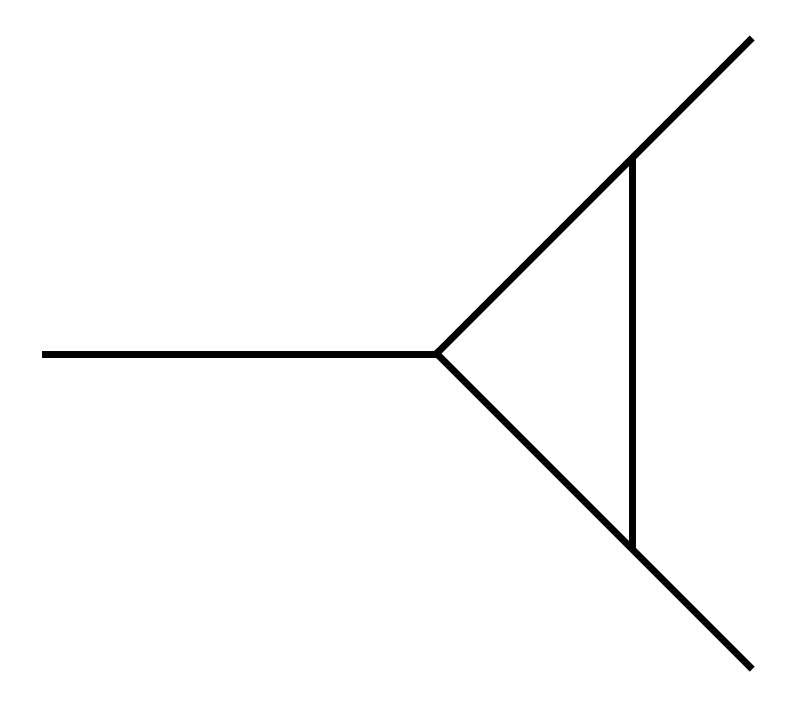}
    \end{subfigure}
    \caption{Example topologies for a leg-type correction (left) and a vertex-type correction (right).}
    \label{fig:example_corrections}
\end{figure}

Finally, for an interference-type process of the kind $(\Phi \to X_3,X_4) \otimes X$, where $X$ labels the spectator particle that forward scatters with a four-momentum $q$,
we get different expressions when the spectator appears from the partial cut of a leg correction loop and when the spectator appears from the partial cut of a vertex correction loop, as shown in Fig.~\ref{fig:example_corrections}. For a leg interference-type process, we get
\begin{equation}
\begin{aligned}
    \frac{\mathrm{d}n_\Phi}{\mathrm{d}t}\eval_{(1 \to 2) \otimes X\mathrm{-leg}} 
    &= \frac{\eta_X}{8(2\pi)^6} \ e^{\mu_X/T}
    \int_{m_\Phi}^\infty \dd{\omega}  
    \int_{E_{3,\mathrm{min}}}^{E_{3,\mathrm{max}}}\dd{E_3} 
    \int_{m_q}^\infty \dd{E_q} 
    \int_{-1}^{1} \dd{\cos\theta} 
    \int_{0}^{2\pi} \dd{\phi} 
    \\&\qquad \qquad
    e^{-\omega/T} e^{-E_q/T} \frac{q}{2E_3^2}
    \qty{
    -\tilde{\M}_L\tilde{\M}_R
    + E_3 \qty[\dv{p_3^0}\tilde{\M}_L\tilde{\M}_R]_{p_3^0=E_3}
    }\,,
\end{aligned}
\end{equation}
where $\eta_X=1,\,-1$ if the spectator is a fermion or a boson respectively, 3 is the label for the particle with the double pole, $\vb{p}_3 \cdot \vb{q} \sim \cos\theta$, and $\tilde{\M}_L\tilde{\M}_R={\M}_L{\M}_R \times (p_3^2 - m_3^2)$ are the matrix elements with the on-shell propagator factored out.
For a vertex interference-type process, we get
\begin{equation}
\begin{aligned}
    \frac{\mathrm{d}n_\Phi}{\mathrm{d}t}\eval_{(1 \to 2) \otimes X\mathrm{-vertex}} 
    &= \frac{\eta_X}{8(2\pi)^6} \ e^{\mu_X/T}
    \int_{m_\Phi}^\infty \dd{\omega}  
    \int_{E_{3,\mathrm{min}}}^{E_{3,\mathrm{max}}}\dd{E_3} 
    \int_{m_q}^\infty \dd{E_q} 
    \int_{-1}^{1} \dd{\cos\theta} 
    \int_{0}^{2\pi} \dd{\phi} 
    \\&\qquad \qquad
    e^{-\omega/T} e^{-E_q/T} q \,
    \qty[{\M}_L{\M}_R]
    \label{eq:interf_vertex}
\end{aligned}
\end{equation}
By using the equations presented above, we next calculate the particle production rate in two toy models using the SSHB approach.

\subsection{All-scalar toy model}
For the all-scalar toy model discussed in the previous sections and given by the Lagrangian in Eq.~\eqref{eq:scalar_lag}, 
the imaginary part of the self energy up to loop order $n=2$ is given by, 
\begin{equation}
    \Im \Pi_\Phi = \mathop{\adjincludegraphics[height=2.3cm, valign=c]{diagrams/Scalar-2loop-one-loop.png}}_{\mathrm{I}} +\mathop{\adjincludegraphics[height=2.3cm, valign=c]{diagrams/Scalar-2loop-sym.png}}_{\mathrm{II}}
    +\mathop{\adjincludegraphics[height=2.3cm, valign=c]{diagrams/Scalar-2loop-asym.png}}_{\mathrm{III}}
    \label{eq:scalar_toy_cuts}
\end{equation}
where we identify $\Phi,\,\phi_1,\,\phi_2,\,\phi_3$ with thick, solid, dashed and dotted lines respectively as in Sec.~\ref{sec:SSHBrules}. We choose the mass-hierarchy,
\begin{equation}
\begin{gathered}
    m_{\phi_1} + m_{\phi_2} < m_\Phi\\
    m_\Phi < m_{\phi_1} + m_{\phi_2} + m_{\phi_3}\\
    m_{\phi_3} < m_{\phi_1} + m_{\phi_2} + m_{\Phi}\,,
\end{gathered}
\label{eq:massHierarchyScalar}
\end{equation}
such that all $1\to 2$ processes (apart from $\Phi\to\phi_1\phi_2$) and all $1\to3$ processes are kinematically forbidden. Implementing the SSHB, the three bisections in Eq.~\eqref{eq:scalar_toy_cuts} result in the following kinematically allowed processes contributing to the production rate,
\begin{align}
    \mathop{\adjincludegraphics[height=2.3cm, valign=c]{diagrams/Scalar-2loop-one-loop.png}}_{\mathrm{I}} &= \mathop{\abs{\adjincludegraphics[height=2.0cm, valign=c]{diagrams/Scalar-acut-0left.png}}^2}
    \label{eq:decay_scalar}\\
    \mathop{\adjincludegraphics[height=2.3cm, valign=c]{diagrams/Scalar-2loop-sym.png}}_{\mathrm{II}}
    &=\mathop{\abs{\adjincludegraphics[height=2.0cm, valign=c]{diagrams/Scalar-cut2to2s-0213.png}}^2}_{\mathrm{II\,(a)}}
    +\mathop{\abs{\adjincludegraphics[height=2.0cm, valign=c]{diagrams/Scalar-cut2to2t-0123.png}}^2}_{\mathrm{II\,(b)}}
    +\mathop{\abs{\adjincludegraphics[height=2.0cm, valign=c]{diagrams/Scalar-cut2to2t-0321.png}}^2}_{\mathrm{II\,(c)}}\label{eq:scatt_scalar}\\
    \mathop{\adjincludegraphics[height=2.3cm, valign=c]{diagrams/Scalar-2loop-asym.png}}_{\mathrm{III}} 
    &=\mathop{ \adjincludegraphics[height=2.0cm, valign=c]{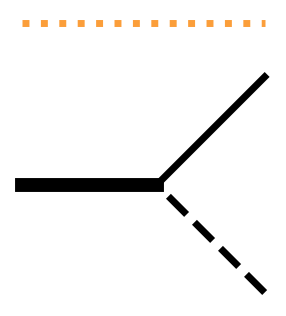}
    \otimes
    \left(\adjincludegraphics[height=2.3cm, valign=c]{diagrams/Scalar-acut-intf-30.png}\right.}_{\mathrm{III\,(a)}}\left.
    +\adjincludegraphics[height=2.3cm, valign=c]{diagrams/Scalar-acut-intf-31.png}\right)\nonumber\\
    &+ \mathop{\adjincludegraphics[height=2.0cm, valign=c]{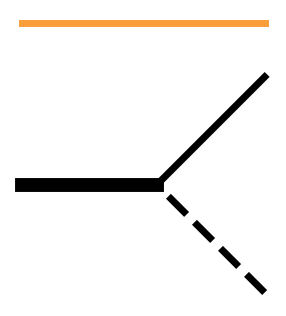}
    \otimes
    \left(\adjincludegraphics[height=2.3cm, valign=c]{diagrams/Scalar-acut-intf-10.png}\right.}_{\mathrm{III\,(b)}}\left.
    +\adjincludegraphics[height=2.3cm, valign=c]{diagrams/Scalar-acut-intf-11.png}\right)\,\label{eq:intf_scalar}.
\end{align}
where the orange lines denote the spectator particles, and ``$+$'' denotes that the associated propagator is on-shell.
The matrix elements for these different processes are given in \App{app:all_scalar_matrix}.

\begin{figure}[t]
    \centering
    \includegraphics[width=0.8\linewidth]{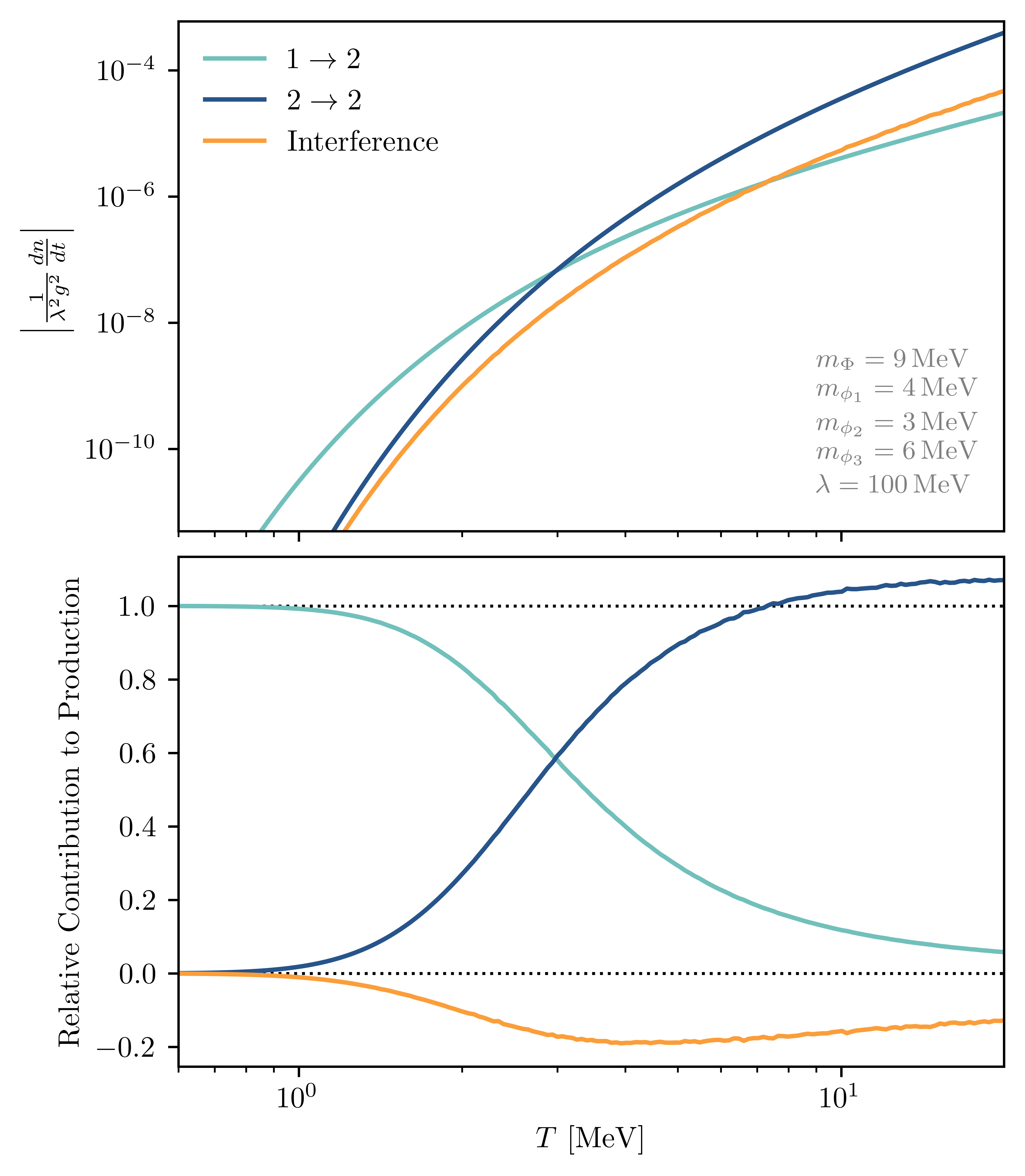}
    \caption{The absolute (top) and relative (bottom) contribution to the production rate per unit volume, $dn/dt$, normalised to the couplings as a function of temperature from different types of processes: $1\rightarrow 2$ decay (\Eq{eq:decay_scalar}), $2\rightarrow 2$ scattering (\Eq{eq:scatt_scalar}), and interference (\Eq{eq:intf_scalar}) for a fixed value of all masses and couplings, and for zero chemical potential, $\mu_i=0$. The interference terms contribute negatively thereby suppressing production. Note that while the relative size of these terms depends on $\lambda$, all processes are proportional to $g^2$, hence the trend is the same for any value of $g$.
    }\label{fig:scalar_ratio_each_term}
\end{figure}

\begin{figure}[t]
    \centering
    \includegraphics[width=0.8\linewidth]{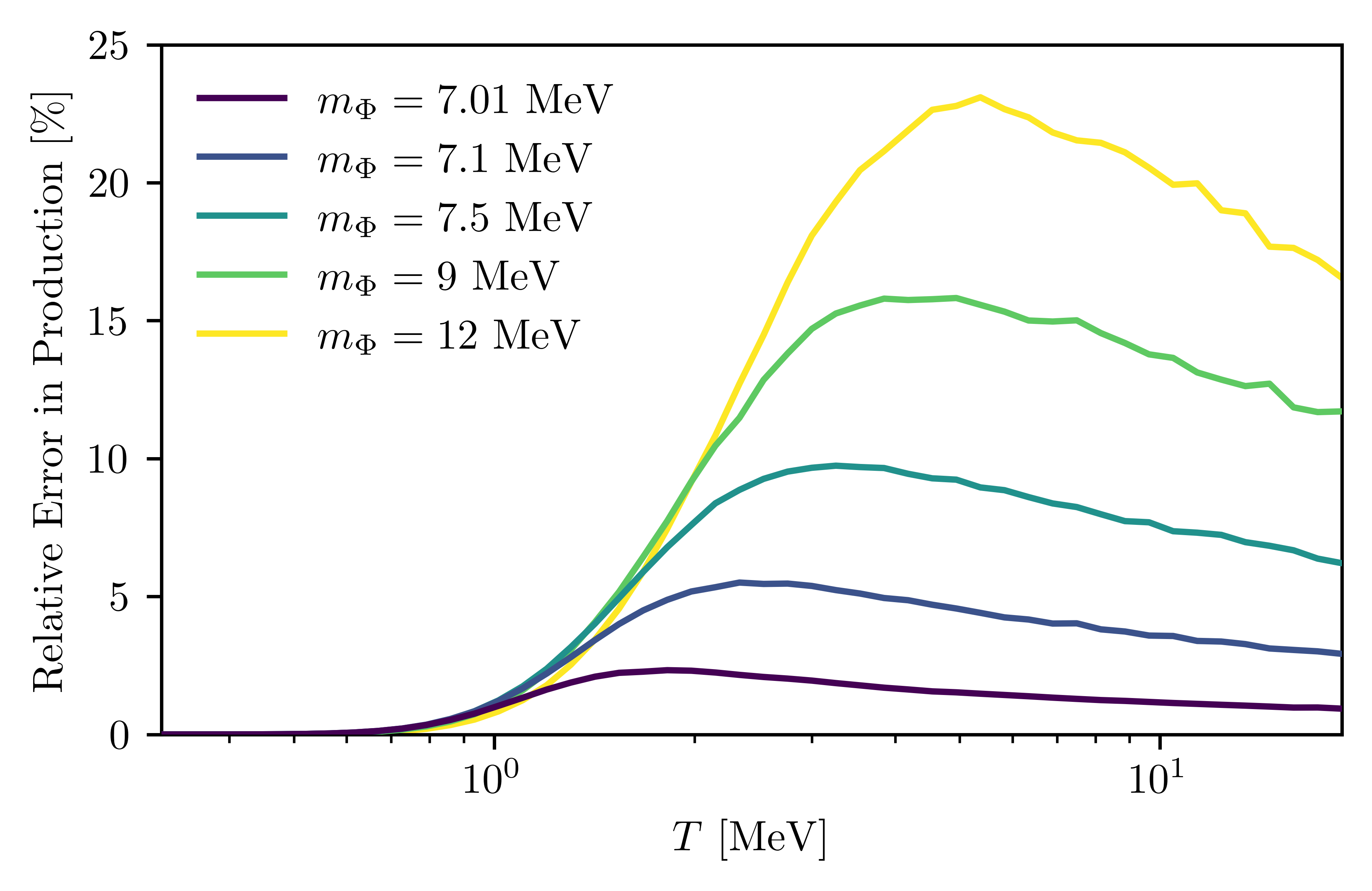}
    \caption{
    Relative
    error in production rate per unit volume as a function of temperature for different values of $m_\Phi$ and with all other parameters fixed as in Fig.~\ref{fig:scalar_ratio_each_term}.}
    \label{fig:scalar_ratio_sums}
\end{figure}

Using Eqs.~\eqref{eq:1to2simple}---\eqref{eq:interf_vertex}, we can now calculate the production rate per unit volume through decays ($1\rightarrow 2$), scattering ($2\rightarrow 2$), and interference. Note that the interference contribution can actually be thought of as a correction to a decay. However, we make the distinction between interference diagrams and decays in order to isolate the effect of interference diagrams on the overall production rate, as shown in Fig.~\ref{fig:scalar_ratio_each_term}.
The scattering and interference terms are dominant at high temperatures, with the interference contribution suppressing the net production rate by $\sim 20\%$ for the parameters shown. 
However, since the scattering and interference contributions are directly proportional to the product of the $\Phi$ and $\phi_i$ phase space densities, they
become Boltzmann suppressed at low temperatures.
  
In Fig.~\ref{fig:scalar_ratio_sums}, we quantify the error in particle production rate as a result of ignoring the interference terms. Specifically, we show the relative difference between
the full two-loop contribution from \Eq{eq:scalar_toy_cuts} and the usual terms involved in the Boltzmann treatment (terms I and II from \Eq{eq:scalar_toy_cuts}). 
The relative error depends on the choice of the coupling $\lambda$ and mass spectrum, necessitating a full parameter scan for models of interest. However, at one fixed value of the parameters in this toy model, we can see that the error can be as large as $\mathcal{O}(1)$ and find other trends. 
Notably, for a given value of $m_\Phi$, the effect of the interference diagrams is greatest for $T\sim m_\Phi$, which is the temperature relevant for freeze-in production \cite{Chu:2011be, Heeba:2023bik, Bringmann:2021sth}. Moreover, the error is largest for the largest kinematically allowed values of $m_\Phi$ with the rest of the mass spectrum left fixed. We expect similar trends to hold for other models and provide an example involving fermions in the following Subsection. 

\subsection{Scalar-fermion-vector theory}
As pointed out in Sec.~\ref{sec:SSHBrules}, the interference terms are sensitive to the signs of the scattering amplitude. Consequently, one has to be careful when considering a medium with background fermions since scattering amplitudes composed of particles may interfere destructively with those including anti-particles. 
To elucidate this effect, we consider a model with a scalar $\Phi$ as the external particle, interacting with a fermion $f$. The fermions are kept in equilibrium through vector interactions mediated by a massive boson $\gamma'$, analogous to a dark photon with, for example, a mass $m_{\gamma^\prime}$ generated by the Stueckelberg mechanism \cite{Stueckelberg:1938hvi}.
The interaction Lagrangian is
\begin{equation}
    \mathcal{L}_\text{int} = g \Phi \bar{f} f - e A_\mu' \bar{f} \gamma^\mu f
    \,.
\end{equation}
For simplicity, we assume a mass hierarchy similar to the all-scalar case in the previous section,
\begin{equation}
\begin{gathered}
    2m_{f} < m_\Phi\\
    m_\Phi < 2m_{f} + m_{\gamma'}\\
    m_{\gamma'} < 2m_{f} + m_{\Phi}
\end{gathered}
\end{equation}
which kinematically forbids several of the $1\leftrightarrow2$ and all of the $1\leftrightarrow3$ processes.
We also assume that the $\Phi$ phase space density is low (as in a freeze-in scenario for example) such that we can neglect contributions from diagrams including internal $\Phi$ particles.
With these assumptions, the imaginary part of the self-energy up to $n=2$ loop order is,
\begin{align}
    \Im &\Pi_\Phi 
    = \mathop{\adjincludegraphics[height=2.1cm, valign=c]{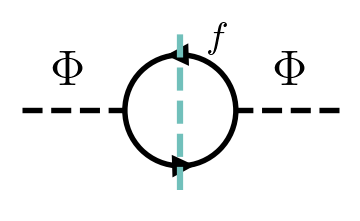}}_{\mathrm{I}}\\
    +&\mathop{
    \adjincludegraphics[height=2.1cm, valign=c]{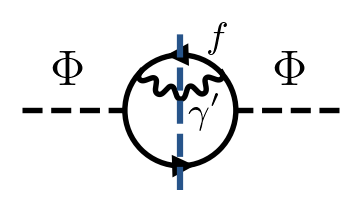}
    +\adjincludegraphics[height=2.1cm, valign=c]{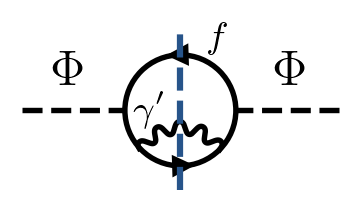}
    +\adjincludegraphics[height=2.1cm, valign=c]{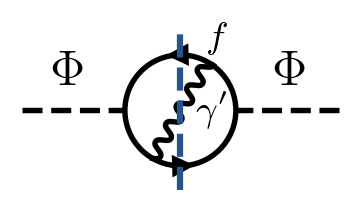}}_{\mathrm{II}} \label{eq:tu_interfere}\\
    +&\mathop{
    \adjincludegraphics[height=2.1cm, valign=c]{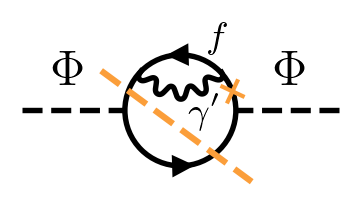}
    +\adjincludegraphics[height=2.1cm, valign=c]{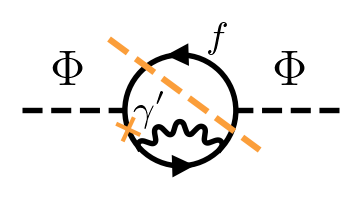}
    +\adjincludegraphics[height=2.1cm, valign=c]{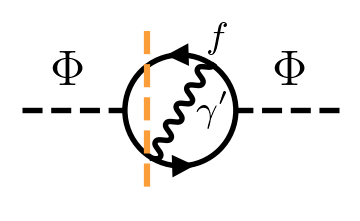}
    +\adjincludegraphics[height=2.1cm, valign=c]{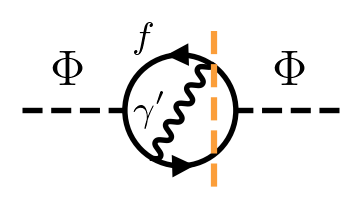}}_{\mathrm{III}}
\end{align}
with
\begin{align}\label{eq:svf_decay}
    \mathop{\adjincludegraphics[height=2.3cm, valign=c]{diagrams/Fermion_sym_one-loop.png}}_{\mathrm{I}} 
    &= \abs{
    \adjincludegraphics[height=2.3cm, valign=c]{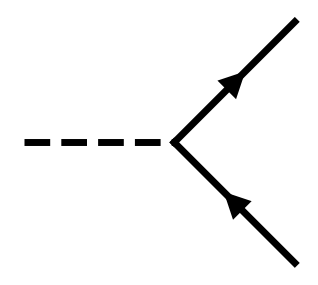}
    }^2
\end{align}
\begin{align}\label{eq:svf_scatt}
    \mathop{\adjincludegraphics[height=2.3cm, valign=c]{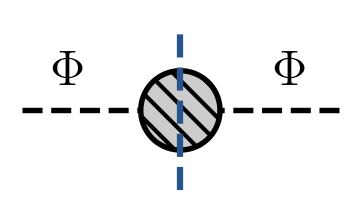}}_{\mathrm{II}} 
    &=\mathop{\abs{
    \adjincludegraphics[height=2.3cm, valign=c]{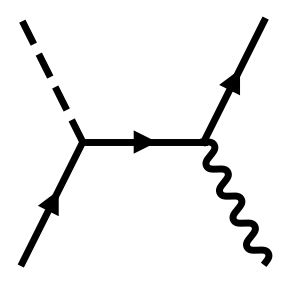}+
    \adjincludegraphics[height=2.3cm, valign=c]{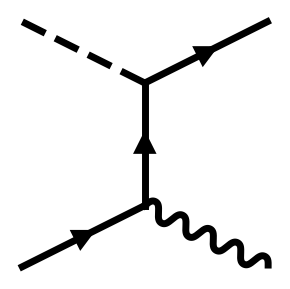}
    }^2}_{\mathrm{II(c)}}\nonumber
    + \mathop{\abs{
    \adjincludegraphics[height=2.3cm, valign=c]{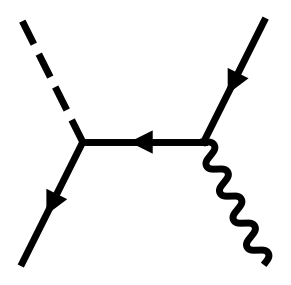}+
    \adjincludegraphics[height=2.3cm, valign=c]{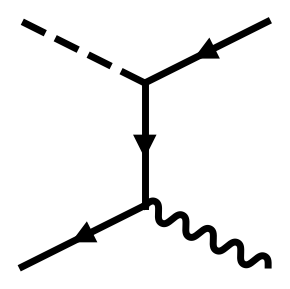}
    }^2}_{\mathrm{II(\bar{c})}}\nonumber\\
    &+ \mathop{\abs{
    \adjincludegraphics[height=2.3cm, valign=c]{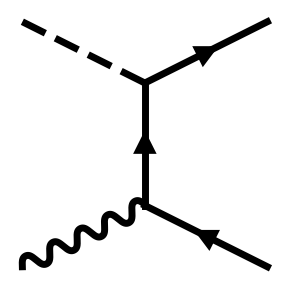}+
    \adjincludegraphics[height=2.3cm, valign=c]{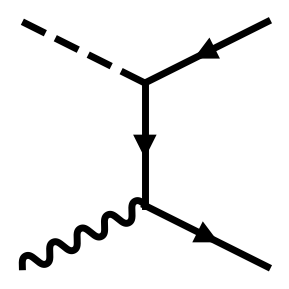}
    }^2}_{\mathrm{II(a)}}
\end{align}
\begin{align}\label{eq:svf_intf}
    \mathop{\adjincludegraphics[height=2.7cm, valign=c]{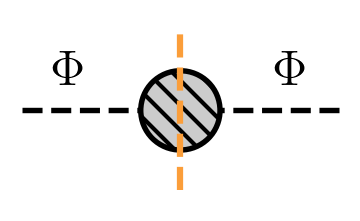}}_{\mathrm{III}}=& \nonumber\\
    % Interference with photon
    &\hspace{-3.5cm}\mathop{
    \adjincludegraphics[height=2.1cm, valign=c]{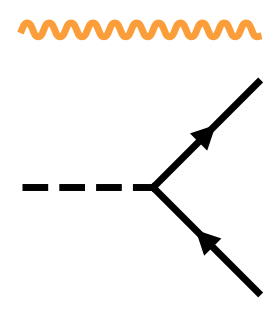}
    \otimes \left(
    \adjincludegraphics[height=2.1cm, valign=c]{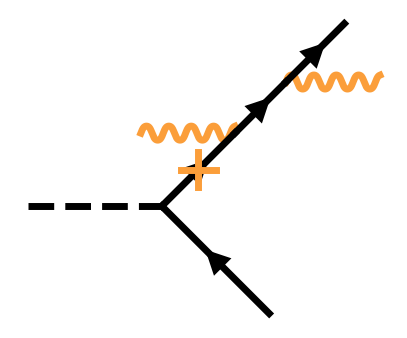}+\right.}_{\mathrm{III(b)-leg}}\left.
    \adjincludegraphics[height=2.1cm, valign=c]{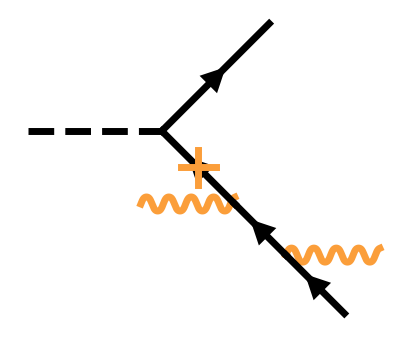}+
    \adjincludegraphics[height=2.1cm, valign=c]{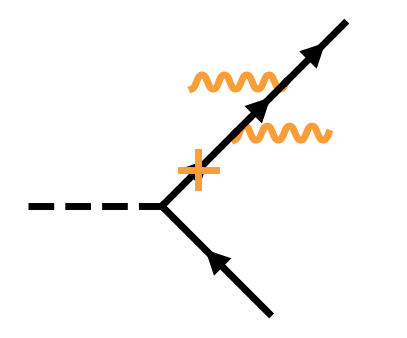}+
    \adjincludegraphics[height=2.1cm, valign=c]{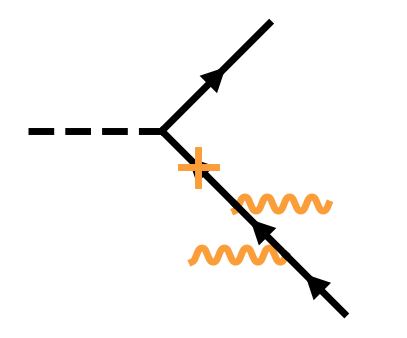}
    \right)^{*}\nonumber\\
    &\hspace{-3.5cm}\mathop{
    +\adjincludegraphics[height=2.1cm, valign=c]{diagrams/Fermion-acut-spec-bf.png}
    \otimes \left(
    \adjincludegraphics[height=2.1cm, valign=c]{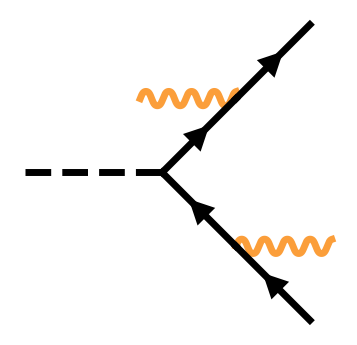}+\right.}_{\mathrm{III(b)-vertex}}\left.
    \adjincludegraphics[height=2.1cm, valign=c]{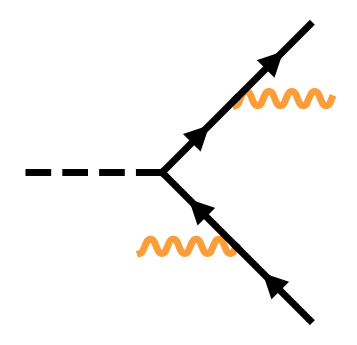}
    \right)^{*} + \qq{c.c.}
    \nonumber\\
    % Interference with fermion
    &\hspace{-3.5cm}\mathop{
    +\adjincludegraphics[height=2.1cm, valign=c]{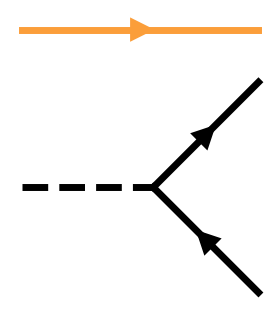}
    \otimes \left(
    \adjincludegraphics[height=2.1cm, valign=c]{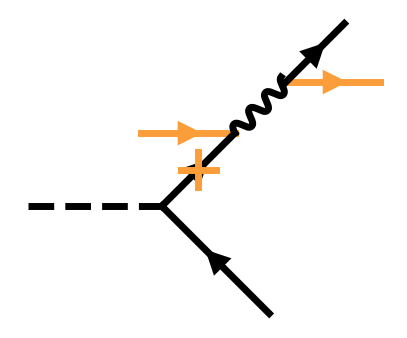}+\right.}_{\mathrm{III(f)-leg}}\left.
    \adjincludegraphics[height=2.1cm, valign=c]{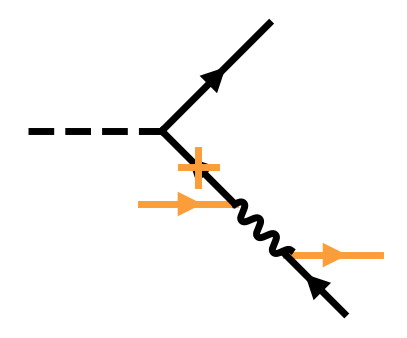}+
    \adjincludegraphics[height=2.1cm, valign=c]{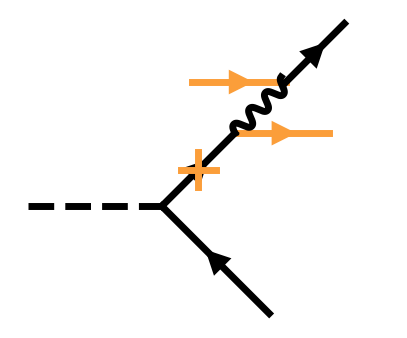}+
    \adjincludegraphics[height=2.1cm, valign=c]{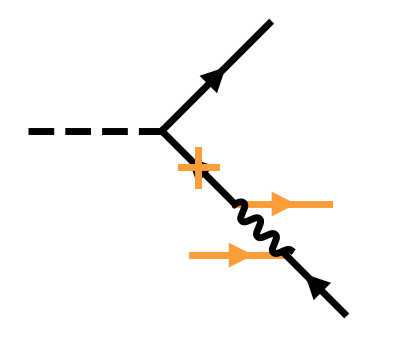}
    \right)^{*}\nonumber\\
    &\hspace{-3.5cm}\mathop{
    +\adjincludegraphics[height=2.1cm, valign=c]{diagrams/Fermion-acut-spec-ff.png}
    \otimes \left(
    \adjincludegraphics[height=2.1cm, valign=c]{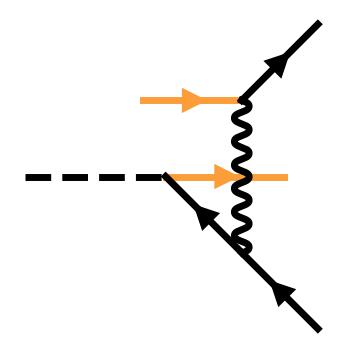}+\right.}_{\mathrm{III(f)-vertex}}\left.
    \adjincludegraphics[height=2.1cm, valign=c]{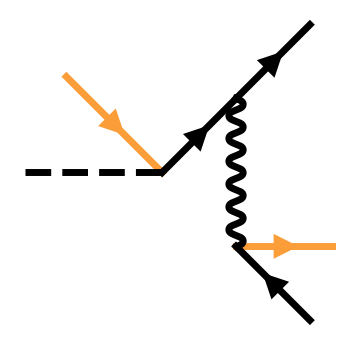}
    \right)^{*} + \qq{c.c.}
    \nonumber\\
    % Interference with antifermion
    &\hspace{-3.5cm}\mathop{
    +\adjincludegraphics[height=2.1cm, valign=c]{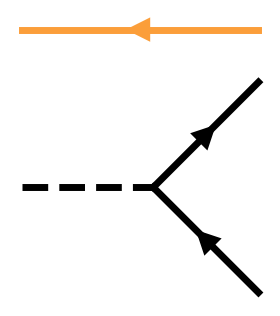}
    \otimes \left(
    \adjincludegraphics[height=2.1cm, valign=c]{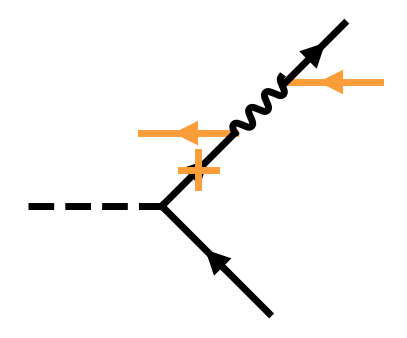}+\right.}_{\mathrm{III(\bar{f})-leg}}\left.
    \adjincludegraphics[height=2.1cm, valign=c]{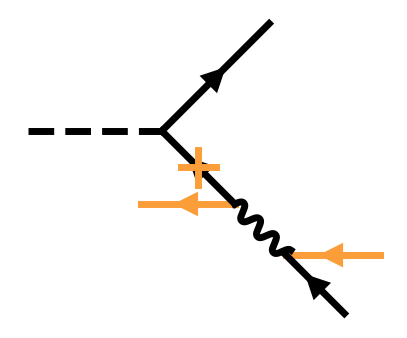}+
    \adjincludegraphics[height=2.1cm, valign=c]{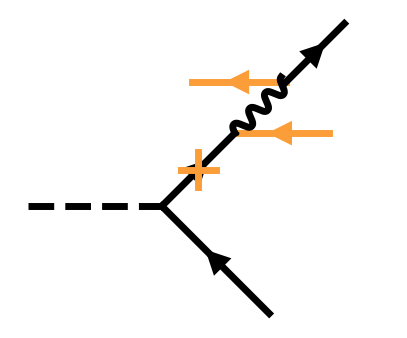}+
    \adjincludegraphics[height=2.1cm, valign=c]{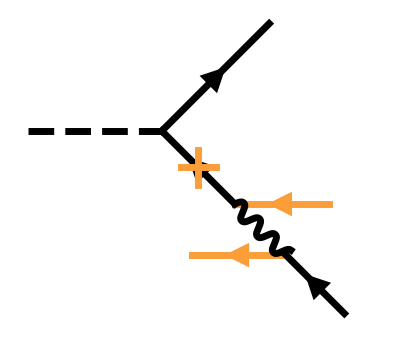}
    \right)^{*}\nonumber\\
    &\hspace{-3.5cm}\mathop{
    +\adjincludegraphics[height=2.1cm, valign=c]{diagrams/Fermion-acut-spec-fbf.png}
    \otimes \left(
    \adjincludegraphics[height=2.1cm, valign=c]{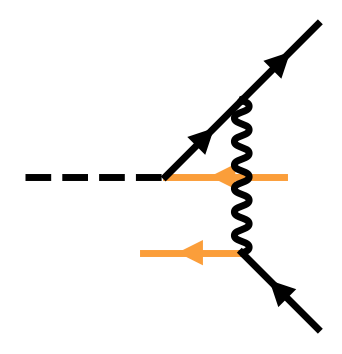}+\right.}_{\mathrm{III(\bar{f})-vertex}}\left.
    \adjincludegraphics[height=2.1cm, valign=c]{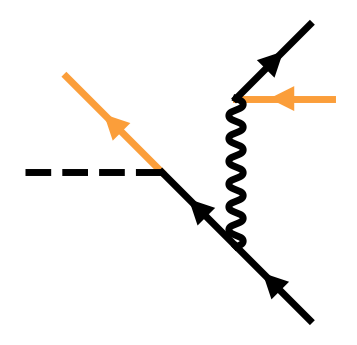}
    \right)^{*} + \qq{c.c.}
\end{align}
In the equations above, we distinguish between leg-type and vertex-type corrections coming from the asymmetric bisection III (see also Fig.~\ref{fig:example_corrections}). The matrix elements for these processes are provided in \App{app:sfv_matrix}. 

The use of SSHB for this toy model illustrates a few of the subtleties pointed out in Sec.~\ref{sec:SSHBrules}:
\begin{itemize}
    \item  We note that the cut corresponding to the third  diagram in Eq.~\eqref{eq:tu_interfere} arising out of the symmetric bisection II, gives rise to interference terms between $s$-channel and $t$-channel amplitudes (or between $t$-channel and $u$-channel amplitudes, depending on the process). These terms conspire with those from the other diagrams arising out of bisection II to yield the rate of $2\to2$ processes with vacuum analogues. These terms are different from the interference terms arising from the asymmetric bisection III which have no vacuum analogues.
    \item The amplitude of the interference terms from bisection III have different signs depending on whether they arise from a leg-type or a vertex-type correction. As a result, different interference diagrams can either enhance or suppress production. In this model, the total interference contribution turns out to be negative.
    \item The $t$-channel diagrams arising out of bisection II have collinear singularities (that exist even in vacuum), which are exactly canceled when one includes the interference contributions at finite temperature. It has been shown previously that in fact both IR and collinear singularities cancel at finite temperature when properly accounting for all physical processes, including the interfering terms, in a medium \cite{Majumder:2001iy}. This cancellation is usually attributed to a generalisation of the Kinoshita$-$Lee$-$Nauenberg theorem to finite temperature (see Refs.~\cite{Kinoshita:1962ur,PhysRev.133.B1549,Altherr:1993tn,LEBELLAC1992423,PhysRevD.44.3955} and references therein).
    \item The SSHB prescription is only valid in the regime where the effective thermal masses of the particles in the loop are smaller than their vacuum masses. In this toy model, one can estimate the thermal masses of the particles in the loop using results derived for QED \cite{1992ApJ...392...70B,Braaten:1993jw},
    \begin{equation}
\begin{gathered}
    m_\mathrm{th}^2 = 
    \begin{cases}
        \pi \alpha T^2/2 &\qq{fermion}\\
        4 \pi \alpha T^2/9 &\qq{boson}\,
    \end{cases}
\end{gathered}
\label{eq:effective_thermal_mass}
\end{equation}
where $\alpha = e^2/4\pi$. Consequently, the prescription is only valid when $m_i > m_{i,\,\mathrm{th}}$ or for $T\lesssim m_i/e$.

\end{itemize}

\begin{figure}[t]
    \centering
    \includegraphics[width=0.8\linewidth]{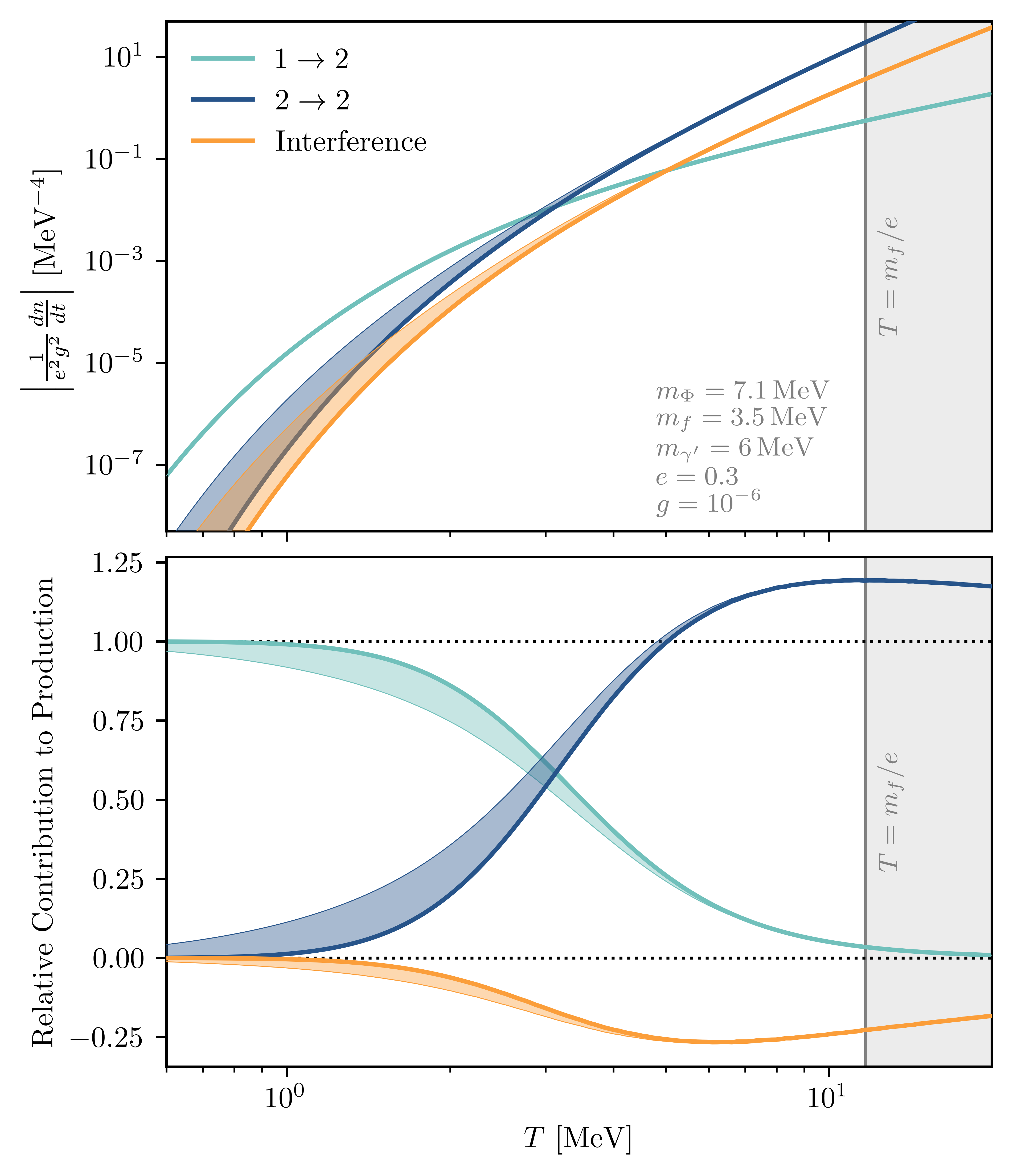}
    \caption{
    The absolute (top) and relative (bottom) contribution to the production rate per unit volume, $dn/dt$, normalised to the couplings as a function of temperature from different types of processes: $1\rightarrow 2$ decay (\Eq{eq:svf_decay}), $2\rightarrow 2$ scattering (\Eq{eq:svf_scatt}) and interference (\Eq{eq:svf_intf}) for a fixed value of all masses and couplings. We vary the chemical potential, $\mu_f \in [0,3]\,\mathrm{MeV}$ resulting in the band between the thick and thin lines respectively. The shaded region with $T>m_f/e$ denotes the temperature for which the thermal mass corrections to the interacting particles cannot be ignored and the formalism needs to be extended (see text for details).
    }
    \label{fig:fermion_ratio_each_term}
\end{figure}
\begin{figure}[t]
    \centering
    \includegraphics[width=0.8\linewidth]{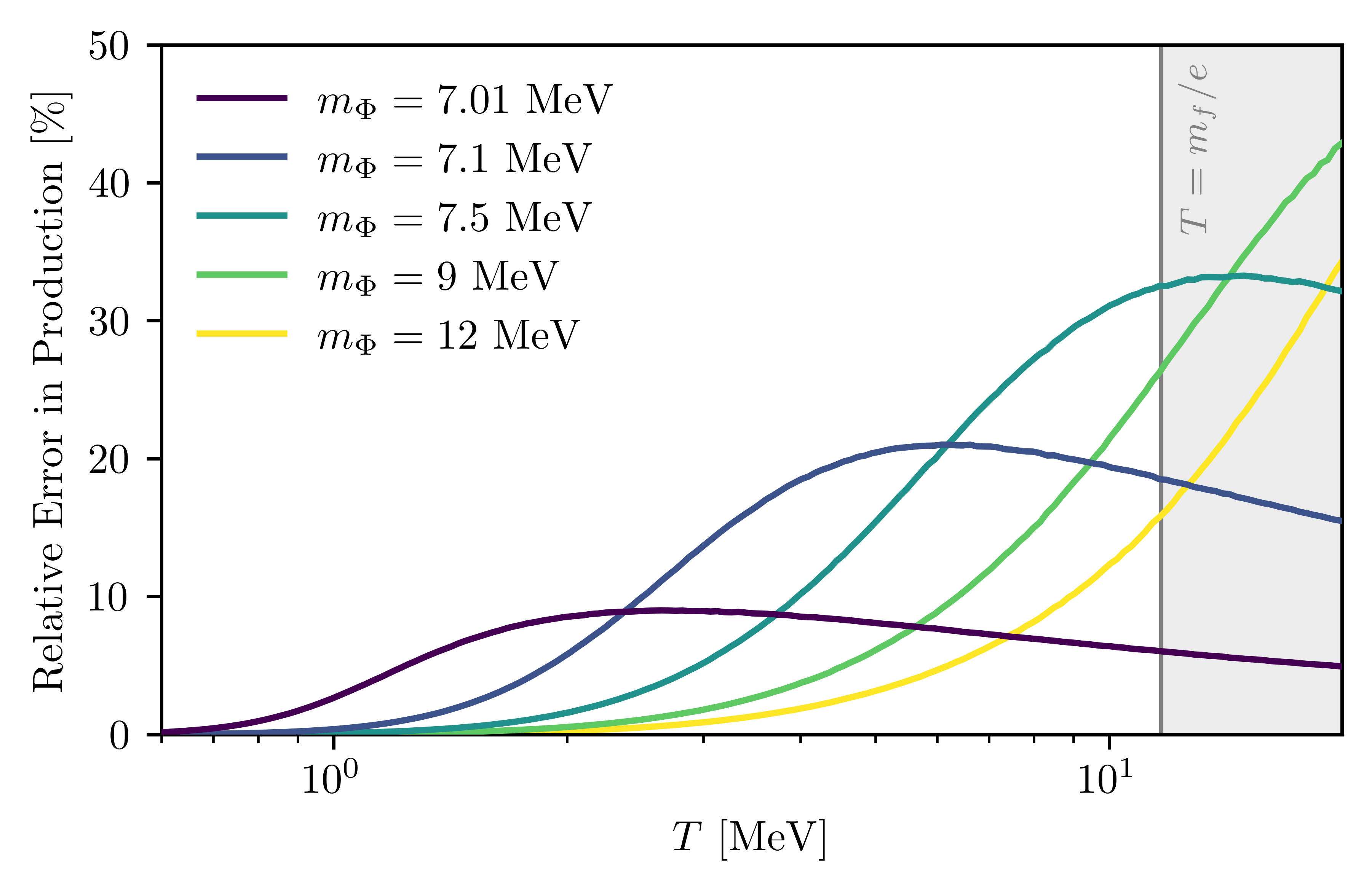}
    \caption{Relative
    error in the production rate per unit volume as a function of temperature for different values of $m_\Phi$, with all other parameters fixed as in Fig.~\ref{fig:fermion_ratio_each_term}. The shaded high-$T$ region is as in Fig.~\ref{fig:fermion_ratio_each_term}.}
    \label{fig:fermion_ratio_sums}
\end{figure}
\begin{figure}[t]
    \centering
    \includegraphics[width=0.8\linewidth]{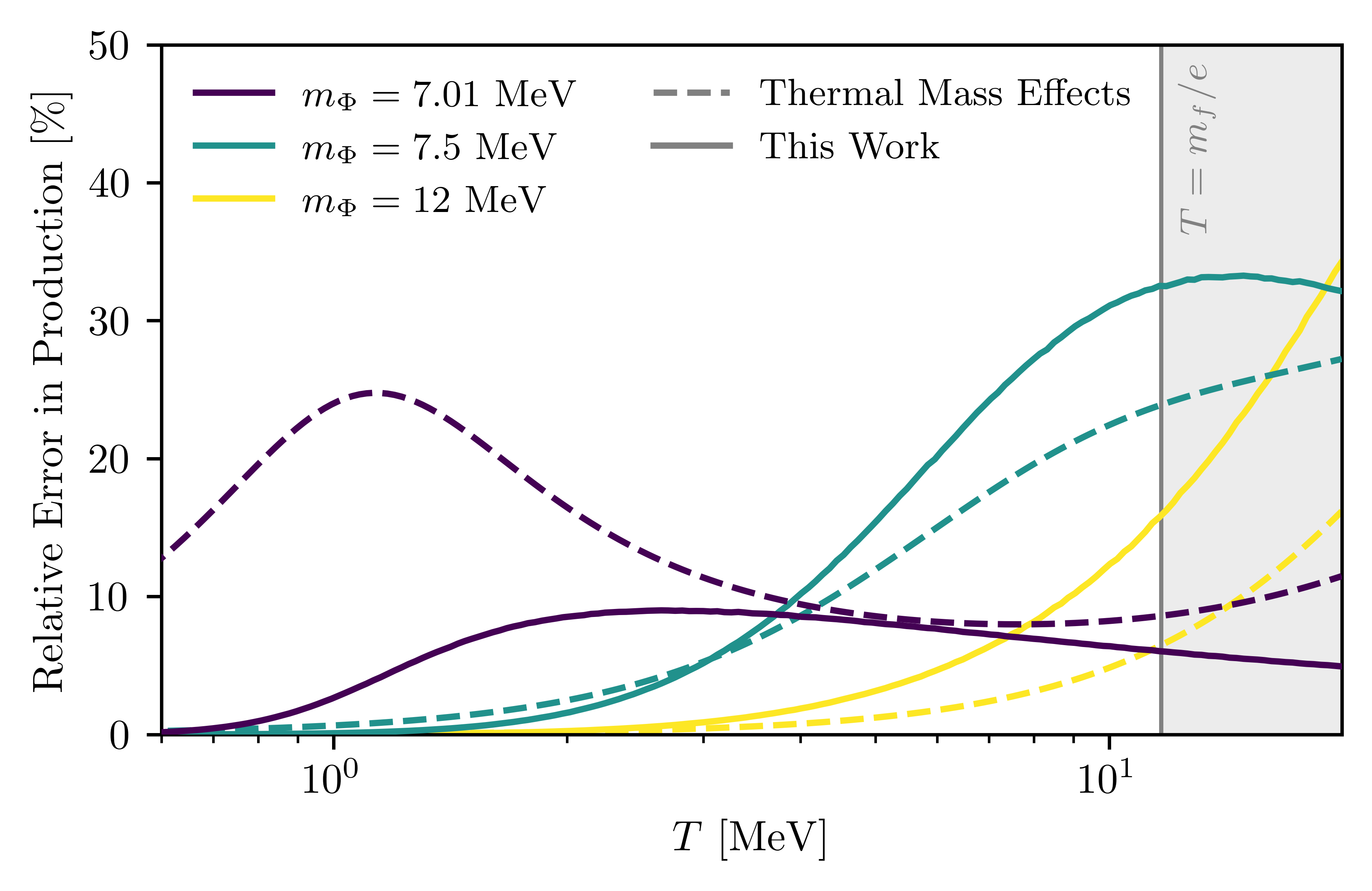}
    \caption{
    Similar to figure \Fig{fig:fermion_ratio_sums} but illustrating the relative error in production when ignoring the thermal masses of the particles and when ignoring the interference contributions highlighted in this work. The shaded high-$T$ region is as in Fig.~\ref{fig:fermion_ratio_each_term}.}
    \label{fig:fermion_ratio_sums_thermalmass}
\end{figure}

In Fig.~\ref{fig:fermion_ratio_each_term}, we 
show the relative contributions of different kinds of diagrams to the overall production rate, with the gray solid line demarcating the range of validity of our prescription. 
As in the scalar case, the net effect of the interference terms is to suppress the total production rate. However, individual interference terms can be either positive or negative (see \App{app:sfv_matrix}). Additionally, we quantify the impact of finite density on this model by including the fermion chemical potential $\mu_f$. Varying $\mu_f$ between 0 and 3~MeV, with the upper limit ensuring that we are always in the non-degenerate limit, produces the band bounded by the thick and thin lines in Fig.~\ref{fig:fermion_ratio_each_term}.
Since the combination of phase space factors in the $1 \to 2$ process remains invariant under the exchange of fermions and anti-fermions, it is independent of the chemical potential.
In contrast, this symmetry does not hold for the $2 \to 2$ and interference diagrams, where the production rate exhibits a clear dependence on $\mu_f$, with the effect being most significant when $T < \mu_f$. 
Overall, we find that for moderately large values of $\mu_f \sim m_f$, where the fermions are not yet fully degenerate, the influence of the chemical potential on the production rate is a few percent.

In Figs.~\ref{fig:fermion_ratio_sums} and~\ref{fig:fermion_ratio_sums_thermalmass}, we quantify the error in the production rate arising from omitting the interference terms. Fig.~\ref{fig:fermion_ratio_sums} shows the error for a range of $\Phi$ masses with other parameters held fixed. As in the scalar case, the interference contributions are substantial for $T\gtrsim m_\Phi$, hinting at potentially significant corrections to the freeze-in dark matter predictions. Meanwhile, Fig.~\ref{fig:fermion_ratio_sums_thermalmass} shows a comparison between the error arising from omitting the interference terms and the error arising from neglecting thermal mass corrections.
The latter constitutes one of the primary FTD effects typically considered in DM production studies, and has been shown to significantly alter the observables \cite{Bringmann:2021sth, Heeba:2019jho, Lebedev:2019ton,Darme:2019wpd,Dvorkin:2019zdi}. Thermal mass effects are often incorporated in a somewhat heuristic way by adding a correction to the vacuum mass appearing in the rates \cite{No:2019gvl, Chakrabarty:2022bcn}, 
\begin{equation}
    m^2 \to m_\mathrm{eff}^2(T) = m^2 + m_\mathrm{th}^2(T).
\end{equation}
Here, we estimate the thermal masses of the particles running in the loop by using the functional form of SM thermal masses in a QED plasma from Eq.~\ref{eq:effective_thermal_mass}.

As shown in Fig.~\ref{fig:fermion_ratio_sums_thermalmass}, the relative contributions to the net particle production rate are similar when comparing interference diagrams to thermal mass effects. This further highlights the need for these interference diagrams to be included in calculations requiring precision better than $\mathcal{O}(1)$.

\section{Conclusions and outlook}\label{sec:conclusion}
Many astrophysical and cosmological tests of BSM physics involve the production of weakly coupled particles from, or their interaction with, finite-density and finite-temperature plasmas. The presence of this ambient background substantially modifies these processes, and therefore the associated observables. In this work, we present a complete set of rules to calculate in-medium production rates of BSM particles by reducing particle self-energies in the ITF at arbitrary loop order $n$ to a sum over tree-level processes. The latter arises as a result of all possible bisections, or cuts, of the self-energy diagrams. Symmetric bisections, where the two halves of the bisected self-energy have no internal loops, give rise to the rates typically used in the BSM literature~\cite{Weldon:1983jn}. 
However, a self-consistent calculation must include contributions from asymmetric bisections, where one or both halves of the bisected self-energy have internal loops. These kinds of diagrams necessarily generate an interference between scattering amplitudes for
$n\to m$ processes (where other background fields are simply ``spectators'') and scattering amplitudes for $n+\ell \to m+ \ell$ processes (where $\ell$ background fields forward scatter without transferring any momentum). The interference between these diagrams is a completely in-medium effect with no vacuum analog.
Since these interference contributions are expressed as a product of amplitudes (corresponding to the two halves of the bisection)
, the individual signs of the amplitudes play a crucial role in determining whether the overall contribution enhances or suppresses the particle production rate. 
Additionally, as has been shown in other contexts~\cite{Majumder:2001iy, LEBELLAC1992423, PhysRevD.44.3955}, these terms are also necessary to regulate IR and collinear divergences at FTD.  

We applied the SSHB rules presented in this work to two toy models to quantify the impact of these interference contributions on BSM particle production, and found that these can modify particle production by an $\mathcal{O}(1)$ amount. Interestingly, this modification is maximal at temperatures similar to the mass of the BSM particle, pointing to potentially significant corrections to the predicted dark matter abundance in a diversity of well-established freeze-in models. The SSHB contributions may also impact observables that rely on the production of light BSM particles from low-temperature plasmas.   

Our formalism is valid in the regime where the thermal masses of the particles in equilibrium can be ignored. However, we note  that the SSHB formalism still holds if the thermal masses can be assumed to be purely functions of the temperature or the chemical potential. In this case, the poles of the effective in-medium propagators only shift by a constant amount, independent of the corresponding particle's four-momentum. This implies that the SSHB prescription can be applied and the in-medium rates can be calculated as detailed in the sections above. An extension of the current formalism to account for momentum-dependent thermal masses, which requires resumming the propagators involved, has been left to future work. We also leave the extension of SSHB to anisotropic plasmas (which involve different forms of propagators among other subtleties, as discussed in Ref.~\cite{Brahma:2024vxb}) to future work. It would also be interesting to further explore the connection between the interference-type diagrams explored here and the cancellation of divergences, as well as to determine whether some version of the SSHB cutting rules can be applied to non-equilibrium systems using the real-time formalism. Additionally, we expect new insights arising through the correspondence between SSHB and related descriptions of condensed matter systems, as was recently highlighted in Ref.~\cite{Liang:2025nan}.

We envision several avenues for the application of SSHB to questions within and beyond the SM. For instance, for on-shell photons, it is well known that the primary damping channels are $\mathcal{O}(\alpha^2)$, corresponding to a two-loop self-energy diagram \cite{1992ApJ...392...70B,Braaten:1993jw}. Generalizing this to off-shell photons, as was done in Ref.~\cite{Scherer:2024cff} which computed only the $\mathcal{O}(\alpha)$ self-energy, will enable accurate FTD calculations for processes involving photons in all parts of phase space. In this work, we explored two relatively simple toy models to illustrate the correspondence between SSHB and a full thermal field theory calculation. Applying SSHB to benchmark BSM models and observables of interest, for instance the freeze-in production of DM and particle emission in stars, will be the subject of future works.

\section*{Acknowledgements}
It is a pleasure to thank Torsten Bringmann, Simon Caron-Huot, Natnael Debru, Charles Gale, and Felix Kahlhoefer for useful discussions pertaining to this work. NB was supported in part by a Doctoral Research Scholarship from the Fonds de Recherche du Qu\'ebec -- Nature et Technologies and by the Canada First Research Excellence Fund through the Arthur B. McDonald Canadian Astroparticle Physics Research Institute. SH acknowledges support from the Canadian Institute of Particle Physics through the IPP 50th Anniversary Connect Fellowship and from the Fonds de Recherche du Qu\'ebec -- Nature et Technologies through the Programme \'Etablissement de la Rel\`eve Professorale. HS acknowledges support from the Natural Sciences and Engineering Research Council of Canada as a Vanier Scholar. NB, SH, HS, and KS acknowledge support from a Natural Sciences and Engineering Research Council of Canada Subatomic Physics Discovery Grant, from the Canada Research Chairs program, and from the CIFAR Global Scholars program. 

The Feynman diagrams in this project were made using \texttt{Feyndraw} \cite{scherer_2025_16103298}. 
The numerical analysis made use of \texttt{FeynArts} \cite{Hahn:2000kx}, \texttt{FeynCalc} \cite{Shtabovenko_2025,Shtabovenko_2020,Shtabovenko_2016,MERTIG1991345}, \texttt{Numpy} \cite{Harris_2020}, \texttt{Scipy} \cite{Virtanen_2020}, \texttt{Matplotlib} \cite{4160265}, and \texttt{Mathematica} \cite{Mathematica}.

\appendix

\section{Identities for higher order poles}
In this appendix we gather all the mathematical identities involving higher order poles that enter in our multi-loop diagrams.

\subsection{Residue identities}
\label{app:res_iden}
For a function $g(p) = g(p_0,\vb{p})$ which has no pole at $p_0 = k_0 \pm E_{p-k}$, we can write the identity \Eq{eq:res_simplify}. The detailed computation to obtain the residue at $p_0 = k_0 + E_{p-k}$ is as follows:
\begin{flalign}
&\int \dkbar[3]{p} \, \text{Res} \bigg\{ [D(p_0-k_0,\vb{p}-\vb{k})]^{n+1} \, g(p) \bigg\}_{p_0 = k_0 + E_{p-k}} \notag \\
&= \int \dkbar[3]{p} \, \frac{1}{n!} \frac{d^n}{dp_0^n} \left[ \frac{g(p)}{(p_0 - k_0 + E_{p-k})^{n+1}} \right]_{p_0 = k_0 + E_{p-k}} \notag \\
&= \int \dkbar[4]{p} \, \frac{1}{n!} \deltabar(p_0 - k_0 - E_{p-k}) \frac{d^n}{dp_0^n} \left[ \frac{g(p)}{(p_0 - k_0 + E_{p-k})^{n+1}} \right] \notag \\
&= \int \dkbar[4]{p} \, \frac{(-1)^n}{n!} \deltabar^{(n)}(p_0 - k_0 - E_{p-k}) \frac{g(p)}{(p_0 - k_0 + E_{p-k})^{n+1}} \notag \\
&= \int \dkbar[4]{p} \, \deltabar(p_0 - k_0 - E_{p-k}) \frac{1}{(p_0 - k_0 - E_{p-k})^n (p_0 - k_0 + E_{p-k})^n} \frac{g(p)}{p_0 - k_0 + E_{p-k}} \notag \\
&= \int \dkbar[4]{p} \, \frac{1}{2 E_{p-k}} \deltabar(p_0 - k_0 - E_{p-k}) \frac{g(p)}{((p_0 - k_0)^2 - E_{p-k}^2)^n} \notag \\
&= \int \dkbar[4]{p} \dkbar[4]{q} \, \deltabar^{4}(q-p+k) \frac{1}{2 E_q} \deltabar(q_0 - E_q) \frac{1}{(q_0^2 - E_q^2)^n} g(p) \notag \\
&= \int \dkbar[4]{p} \dkbar[4]{q} \, \deltabar^{4}(q-p+k) \deltabar^{(+)} (q_0^2 - E_q^2) \frac{1}{(q_0^2 - E_q^2)^n} g(p) \notag \\
&= \int \dkbar[4]{p} \dkbar[4]{q} \, \deltabar^{4}(q-p+k) \frac{(-1)^n}{n!} \deltabar^{(+)(n)}(q_0^2 - E_q^2) g(p) \notag \\
&= \int \dkbar[4]{q} \, \frac{(-1)^n}{n!} \deltabar^{(+)(n)}(q_0^2 - E_q^2) g(p = k + q)
\end{flalign}
The residue at $p_0 = k_0 - E_{p-k}$ is straightforward to obtain in a similar way.

\subsection{Discontinuity identities}
\label{app:discon_iden}
The final step for computing the imaginary part of a self-energy is to take the discontinuity. For each propagator in the integrand that still involves the external energy $\omega$, the discontinuity for $\Pi \sim \qty[D(\omega-p_0,\vb{k}-\vb{p})]^{n+1}$ is given by \Eq{eq:disc_simplify_identity}. The rest of the integrand for the self-energy must have no pole at $\omega = p_0 \pm E_{k-p}$ other than the pole from the propagator in question. The proof of that replacement rule goes as follows:
\begin{equation}
\Pi \sim \frac{1}{[(\omega - p_0)^2 - E_{k-p}^2]^{n+1}} = \frac{1}{[\omega - p_0 - E_{k-p}]^{n+1} [\omega - p_0 + E_{k-p}]^{n+1}}
\end{equation}
so
\begin{multline}
\text{Disc } \Pi \sim 
\frac{1}{[\omega - p_0 - E_{k-p} + i\epsilon]^{n+1} [\omega - p_0 + E_{k-p} + i\epsilon]^{n+1}}
\\- \frac{1}{[\omega - p_0 - E_{k-p} - i\epsilon]^{n+1} [\omega - p_0 + E_{k-p} - i\epsilon]^{n+1}} 
\end{multline}
Define $x_\pm = \omega - p_0 \pm E_{k-p}$, then
\begin{equation}
\begin{aligned}
\Disc{\Pi} &\sim \Bigg[
\mathcal{P} \frac{1}{x_-^{n+1}} - i\pi \frac{1}{x_-^n} \delta(x_-) 
\Bigg] 
\Bigg[
\mathcal{P} \frac{1}{x_+^{n+1}} - i\pi \frac{1}{x_+^n} \delta(x_+)
\Bigg] \\
&\qquad-\Bigg[
\mathcal{P} \frac{1}{x_-^{n+1}} + i\pi \frac{1}{x_-^n} \delta(x_-) 
\Bigg]
\Bigg[
\mathcal{P} \frac{1}{x_+^{n+1}} + i\pi \frac{1}{x_+^n} \delta(x_+)
\Bigg] \\
&= 
\mathcal{P} \frac{1}{x_-^{n+1}} \mathcal{P} \frac{1}{x_+^{n+1}} 
- i\pi \mathcal{P} \frac{1}{x_-^{n+1}} \frac{1}{x_+^n} \delta(x_+) 
- i\pi \mathcal{P} \frac{1}{x_+^{n+1}} \frac{1}{x_-^n} \delta(x_-)
- \pi^2 \frac{1}{x_-^n} \frac{1}{x_+^n} \delta(x_-) \delta(x_+) \\
&- 
\mathcal{P} \frac{1}{x_-^{n+1}} \mathcal{P} \frac{1}{x_+^{n+1}} 
- i\pi \mathcal{P} \frac{1}{x_-^{n+1}} \frac{1}{x_+^n} \delta(x_+)
- i\pi \mathcal{P} \frac{1}{x_+^{n+1}} \frac{1}{x_-^n} \delta(x_-) 
+ \pi^2 \frac{1}{x_-^n} \frac{1}{x_+^n} \delta(x_-) \delta(x_+)\\
&= - 2\pi i \frac{1}{(x_+ x_-)^n} \qty[\frac{1}{x_-} \delta(x_+) + \frac{1}{x_+} \delta(x_-)]
\end{aligned}
\end{equation}
Finally,
\begin{equation}
\begin{aligned}
\text{Disc }\Pi &\sim -2\pi i \frac{1}{[(\omega - p_0)^2 - E_{k-p}^2]^n} 
\left[ 
\frac{1}{2 E_{k-p}} \delta(\omega - p_0 - E_{k-p})
- \frac{1}{2 E_{k-p}} \delta(\omega - p_0 + E_{k-p}) 
\right] \\
&= -2\pi i \int \frac{\dkbar[4]{q} (2\pi)^4 }{(q_0^2 - E_{q}^2)^n 2E_q} 
\left[ \delta^{(4)}(k-p-q) \delta(q_0 - E_{q})
- \delta^{(4)}(k-p+q) \delta(-q_0 + E_{q}) 
\right] \\
&= -2\pi i \int \dkbar[4]{q} \frac{1}{(q_0^2 - E_{q}^2)^n} \delta^{(+)}(q_0^2 - E_{q}^2)
(2\pi)^4 \left[ \delta^{(4)}(k-p-q)
- \delta^{(4)}(k-p+q)
\right] \\
&= - i \int \dkbar[4]{q} (2\pi) \frac{(-1)^n}{n!} \delta^{(+)(n)}(q_0^2 - E_{q}^2)
(2\pi)^4 \left[ \delta^{(4)}(k-p-q)
- \delta^{(4)}(k-p+q)
\right]
\end{aligned}
\end{equation}

\section{Relevant matrix elements}

\subsection{All-scalar theory}\label{app:all_scalar_matrix}
The matrix elements for the different processes in Eqs.~\eqref{eq:decay_scalar}---\eqref{eq:intf_scalar} for the all-scalar model are listed here
\begin{align}
    &\abs{\M}^2_\mathrm{I} = g^2\\
    &\abs{\M}^2_\mathrm{II(a)} = \frac{g^2 \lambda ^2}{(s-m_{\phi_1}^2)^2}\\
    &\abs{\M}^2_\mathrm{II(b)} = \frac{g^2 \lambda ^2}{(t-m_{\phi_1}^2)^2}\\
    &\abs{\M}^2_\mathrm{II(c)} = \frac{g^2 \lambda ^2}{(t-m_{\phi_1}^2)^2}\\
    &[\tilde{\M}_L\tilde{\M}_R]_\mathrm{III(a)}
    = g^2 \lambda ^2 \qty(
    \frac{1}{m_{\phi_3}^2-2 (p_3 \cdot q)}+
    \frac{1}{m_{\phi_3}^2+2 (p_3 \cdot q)})\\
    &[\tilde{\M}_L\tilde{\M}_R]_\mathrm{III(b)}
    = g^2 \lambda ^2 \qty(
    \frac{1}{2m_{\phi_1}^2-m_{\phi_3}^2-2 (p_3 \cdot q)}+
    \frac{1}{2m_{\phi_1}^2-m_{\phi_3}^2+2 (p_3 \cdot q)})
\end{align}
Furthermore,
\begin{align}
    \int_{t_1}^{t_0} \dd{t} \abs{\M}^2_\mathrm{II(a)}
    &= \frac{g^2 \lambda ^2 s \beta (s,m_{\Phi},m_{\phi_2}) \beta (s,m_{\phi_1},m_{\phi_3})}{(s-m_{\phi_1}^2)^2}\\
    \int_{t_1}^{t_0} \dd{t} \abs{\M}^2_\mathrm{II(b)}
    &= \frac{g^2 \lambda ^2 s \beta (s,m_{\Phi},m_{\phi_1}) \beta (s,m_{\phi_2},m_{\phi_3})}{(t_0-m_{\phi_1}^2) (t_1-m_{\phi_1}^2)}\\
    \int_{t_1}^{t_0} \dd{t} \abs{\M}^2_\mathrm{II(c)}
    &= \frac{g^2 \lambda ^2 s \beta (s,m_{\Phi},m_{\phi_3}) \beta (s,m_{\phi_2},m_{\phi_1})}{(t_0-m_{\phi_1}^2) (t_1-m_{\phi_1}^2)}
\end{align}
and
\begin{flalign}
    &\qty[\dv{p_3^0}
    [\tilde{\M}_L\tilde{\M}_R]_\mathrm{III(a)}]_{p_3^0=E3}
    = -g^2 \lambda ^2 \qty(
    \frac{2 E_3+2 E_q}{\qty(m_{\phi_3}^2+2 (p_3 \cdot q))^2}+
    \frac{2 E_3-2 E_q}{\qty(m_{\phi_3}^2-2 (p_3 \cdot q))^2})\\
    &\qty[\dv{p_3^0}
    [\tilde{\M}_L\tilde{\M}_R]_\mathrm{III(b)}]_{p_3^0=E3}
    = -g^2 \lambda ^2 \left(
    \frac{2 E_3+2 E_q}{\qty(2m_{\phi_1}^2-m_{\phi_3}^2+2 (p_3 \cdot q))^2}\right.\nonumber\\
    &\qquad\qquad\qquad\qquad\qquad\qquad\qquad\qquad\qquad\left.+\frac{2 E_3-2 E_q}{\qty(2m_{\phi_1}^2-m_{\phi_3}^2-2 (p_3 \cdot q))^2}\right)
\end{flalign}

\subsection{Scalar-fermion-vector theory}\label{app:sfv_matrix}
The matrix elements for the different processes in Eqs.~\eqref{eq:svf_decay}---\eqref{eq:svf_intf} for the scalar-fermion-vector model are listed here
\begin{align}
    &\abs{\M}^2_\mathrm{I} = 2 g^2 (m_{\Phi}^2-4 m_{f}^2)\\
    &\abs{\M}^2_\mathrm{II(c)} = \abs{\M}^2_\mathrm{II(\bar{c})} = 4 e^2 g^2 \bigg(
    -\frac{(m_{\gamma'}^2+2 m_{f}^2) (4 m_{f}^2-m_{\Phi}^2)}{(m_{f}^2-t)^2}
    -\frac{m_{f}^2-t}{m_{f}^2-s}\nonumber\\
    &\qquad\qquad\qquad\qquad-\frac{-m_{\gamma'}^2 m_{\Phi}^2+2 m_{f}^4+4 m_{f}^2 (m_{\gamma'}^2+s)+2 s (s-m_{\Phi}^2)}{(m_{f}^2-s)^2} \nonumber\\
    &\qquad\qquad\qquad\qquad-\frac{9 m_{f}^4+2 m_{f}^2 (4 m_{\gamma'}^2-5 m_{\Phi}^2+3 s)+2 m_{\Phi}^4-2 m_{\Phi}^2 s+s^2}{(m_{f}^2-s) (m_{f}^2-t)}
    \bigg)\\
    &\abs{\M}^2_\mathrm{II(a)} = 4 e^2 g^2 \bigg(
    (m_{\gamma'}^2+2 m_{f}^2) (4 m_{f}^2-m_{\Phi}^2) 
    \qty(\frac{1}{(t-m_{f}^2)^2}+\frac{1}{(u-m_{f}^2)^2})\nonumber\\
    &+\frac{(m_{\gamma'}^4-2 m_{\gamma'}^2 s+16 m_{f}^4+4 m_{f}^2 (4 m_{\gamma'}^2-m_{\Phi}^2-2 s)+m_{\Phi}^4+s^2)}{m_{\gamma'}^2+m_{\Phi}^2-s} \qty(\frac{1}{t-m_{f}^2}+\frac{1}{u-m_{f}^2}) 
    \bigg)
\end{align}
\begin{align}
    &[\tilde{\M}_L\tilde{\M}_R]_\mathrm{III(b)-leg} = 
    \frac{32 e^2 g^2 (4 m_{f}^2-m_{\Phi}^2) (2 (p_3 \cdot q)^2+m_{\gamma'}^2 m_{f}^2)}{m_{\gamma'}^4-4 (p_3 \cdot q)^2}\\
    &[\tilde{\M}_L\tilde{\M}_R]_\mathrm{III(b)-vertex}\nonumber\\
    &=
    \frac{4 e^2 g^2}{m_{\gamma'}^2 (m_{\gamma'}^2-2 (p_3 \cdot q)) (2 (p_1 \cdot q)-2 (p_3 \cdot q)+m_{\gamma'}^2)} \left[4 (m_{\Phi}^2-4 m_{f}^2) ((p_3 \cdot q))^2+4 m_{\gamma'}^2 m_{\Phi}^2 (p_3 \cdot q)\right. \nonumber\\
    &\left.-2 (p_1 \cdot q) ((2 m_{\Phi}^2-8 m_{f}^2) (p_3 \cdot q)+m_{\gamma'}^2 m_{\Phi}^2)+m_{\gamma'}^2 (4 m_{f}^2-m_{\Phi}^2) (3 m_{\gamma'}^2+4 m_{f}^2-2 m_{\Phi}^2)\right]\nonumber\\
    &+\frac{4 e^2 g^2}{m_{\gamma'}^2 (2 p_3 \cdot q+m_{\gamma'}^2) (-2 (p_1 \cdot q)+2 p_3 \cdot q+m_{\gamma'}^2)} \left[4 (m_{\Phi}^2-4 m_{f}^2) ((p_3 \cdot q))^2-4 m_{\gamma'}^2 m_{\Phi}^2 (p_3 \cdot q)\right. \nonumber\\
    &\left.+2 (p_1 \cdot q) ((8 m_{f}^2-2 m_{\Phi}^2) (p_3 \cdot q)+m_{\gamma'}^2 m_{\Phi}^2)+m_{\gamma'}^2 (4 m_{f}^2-m_{\Phi}^2) (3 m_{\gamma'}^2+4 m_{f}^2-2 m_{\Phi}^2)\right]\\
    &[\tilde{\M}_L\tilde{\M}_R]_\mathrm{III(f)-leg} =
    8 e^2 g^2(4 m_{f}^2-m_{\Phi}^2)\qty(\frac{(p_3 \cdot q+2 m_{f}^2)}{2 p_3 \cdot q-m_{\gamma'}^2+2 m_{f}^2}+\frac{(2 m_{f}^2-p_3 \cdot q)}{-2 (p_3 \cdot q)-m_{\gamma'}^2+2 m_{f}^2})
\end{align}
\begin{flalign}
[\tilde{\M}_L\tilde{\M}_R]_\mathrm{III(f)-vertex} &=
    \frac{16 e^2 g^2 ((2 m_{f}^2-m_{\Phi}^2) (p_1 \cdot q)+2 m_{f}^2 (2 p_3 \cdot q+4 m_{f}^2-m_{\Phi}^2))}{(2 p_1 \cdot q+m_{\Phi}^2) (2 p_3 \cdot q-m_{\gamma'}^2 +2 m_{f}^2)}\nonumber\\
    &+\frac{16 e^2 g^2 ((m_{\Phi}^2-2 m_{f}^2) (p_1 \cdot q)+2 m_{f}^2 (-2 (p_3 \cdot q)+4 m_{f}^2-m_{\Phi}^2))}{(m_{\Phi}^2-2 (p_1 \cdot q)) (-2 (p_3 \cdot q)-m_{\gamma'}^2+2 m_{f}^2)}
\end{flalign}
\begin{flalign}
    &[\tilde{\M}_L\tilde{\M}_R]_\mathrm{III(\bar{f})-leg} = [\tilde{\M}_L\tilde{\M}_R]_\mathrm{III(f)-leg}\\
    &[\tilde{\M}_L\tilde{\M}_R]_\mathrm{III(\bar{f})-vertex} = [\tilde{\M}_L\tilde{\M}_R]_\mathrm{III(f)-vertex}
\end{flalign}
and:
\begin{flalign}
    \int_{t_1}^{t_0} \dd{t} \abs{\M}^2_\mathrm{II(c)} &= 2 e^2 g^2 \beta (s,m_{f},m_{\gamma'}) \beta (s,m_{\Phi},m_{f}) \bigg(
    -m_{\gamma'}^2
    -\frac{2 s (m_{\gamma'}^2+2 m_{f}^2) (4 m_{f}^2-m_{\Phi}^2)}{(m_{f}^2-t_0) (m_{f}^2-t_1)}\nonumber\\
    &+\frac{m_{\gamma'}^2 m_{\Phi}^2}{m_{f}^2-s}
    +m_{f}^2
    -\frac{2 s (-m_{\gamma'}^2 m_{\Phi}^2+2 m_{f}^4+4 m_{f}^2 (m_{\gamma'}^2+s)+2 s (s-m_{\Phi}^2))}{(m_{f}^2-s)^2}-
    m_{\Phi}^2+s
    \bigg)\nonumber\\
    &+\frac{4 e^2 g^2 (9 m_{f}^4+2 m_{f}^2 (4 m_{\gamma'}^2-5 m_{\Phi}^2+3 s)+2 m_{\Phi}^4-2 m_{\Phi}^2 s+s^2) \log (\frac{m_{f}^2-t_0}{m_{f}^2-t_1})}{m_{f}^2-s}\\
    \int_{t_1}^{t_0} \dd{t} \abs{\M}^2_\mathrm{II(a)} &=\frac{8 e^2 g^2 s (m_{\gamma'}^2+2 m_{f}^2) (4 m_{f}^2-m_{\Phi}^2) \beta (s,m_{f},m_{f}) \beta (s,m_{\Phi},m_{\gamma'})}{(t_0-m_{f}^2) (t_1-m_{f}^2)}\nonumber\\
    &+\frac{8 e^2 g^2 ((m_{\gamma'}^2-s)^2+16 m_{f}^4+4 m_{f}^2 (4 m_{\gamma'}^2-m_{\Phi}^2-2 s)+m_{\Phi}^4) \log (\frac{t_0-m_{f}^2}{t_1-m_{f}^2})}{m_{\gamma'}^2+m_{\Phi}^2-s}
\end{flalign}
and:
\begin{align}
    &\qty[\dv{p_3^0}[\tilde{\M}_L\tilde{\M}_R]_\mathrm{II(b)-leg}]_{p_3^0=E3} 
    = \frac{16 e^2 g^2}{(m_{\gamma'}^5-4 m_{\gamma'} (p_3 \cdot q)^2)^2} \left[(8 (p_3 \cdot q) (E_3 m_{\gamma'}^6 (p_1 \cdot q)\right. \nonumber\\
    &\left.+E_q m_{\gamma'}^4 (m_{\gamma'}^2+2 m_{f}^2) (4 m_{f}^2-m_{\Phi}^2))-32 E_3 m_{\gamma'}^2 (p_1 \cdot q) (p_3 \cdot q)^3\right. \nonumber\\
    &\left.+16 E_3 (4 m_{f}^2-m_{\Phi}^2) (p_3 \cdot q)^4-16 E_3 m_{\gamma'}^2 (-m_{\gamma'}^2 m_{\Phi}^2+4 m_{f}^4+m_{f}^2 (6 m_{\gamma'}^2-m_{\Phi}^2)) (p_3 \cdot q)^2\right. \nonumber\\
    &\left.+E_3 m_{\gamma'}^6 (m_{\gamma'}^2 m_{\Phi}^2-16 m_{f}^4+4 m_{f}^2 (m_{\gamma'}^2+m_{\Phi}^2)))\right]
\end{align}

\begin{align}
    &\qty[\dv{p_3^0}[\tilde{\M}_L\tilde{\M}_R]_\mathrm{III(f)-leg}]_{p_3^0=E3} = 
    \frac{-64 e^2 g^2}{((m_{\gamma'}^2-2 m_{f}^2)^2-4 (p_3 \cdot q)^2)^2} \left[((p_3 \cdot q) (E_3 (m_{\gamma'}^2-2 m_{f}^2)^2 (p_1 \cdot q)\right. \nonumber\\
    &\left.-E_q (4 m_{f}^4-m_{\gamma'}^4) (4 m_{f}^2-m_{\Phi}^2))+2 E_3 (8 m_{f}^4-m_{\gamma'}^2 m_{\Phi}^2) (p_3 \cdot q)^2-4 E_3 (p_1 \cdot q) (p_3 \cdot q)^3\right. \nonumber\\
    &\left.+E_3 m_{f}^2 (2 m_{\gamma'}^2-m_{\Phi}^2) (m_{\gamma'}^2-2 m_{f}^2)^2\right]
\end{align}

\begin{equation}
\qty[\dv{p_3^0}[\tilde{\M}_L\tilde{\M}_R]_\mathrm{III(\bar{f})-leg}]_{p_3^0=E3} = \qty[\dv{p_3^0}[\tilde{\M}_L\tilde{\M}_R]_\mathrm{III(f)-leg}]_{p_3^0=E3}
\end{equation}

\bibliographystyle{unsrtnat}
\bibliography{main}

\end{document}